\pdfoutput=1 
\documentclass[twocolumn,tighten]{aastex631}

\hypersetup{
    colorlinks=true,
    linkcolor=red,
    filecolor=magenta,      
    urlcolor=cyan,
    citecolor=blue,
    }
    
\usepackage[all]{hypcap}

\shortauthors{Gregg et al.}

\begin{document}

\title{Feedback in emerging extragalactic star clusters, FEAST: The relation between 3.3 $\mu$m PAH emission and Star Formation Rate traced by ionized gas in NGC 628 }

\correspondingauthor{Benjamin Gregg}
\email{bagregg@astro.umass.edu}

\author[0000-0003-4910-8939]{Benjamin Gregg}
\affiliation{Department of Astronomy, 
University of Massachusetts, 710 North Pleasant Street, Amherst, MA 01003, USA}

\author[0000-0002-5189-8004]{Daniela Calzetti}
\affiliation{Department of Astronomy, 
University of Massachusetts, 710 North Pleasant Street, Amherst, MA 01003, USA}

\author[0000-0002-8192-8091]{Angela Adamo}
\affiliation{Department of Astronomy, The Oskar Klein Centre, Stockholm University, AlbaNova, SE-10691 Stockholm, Sweden}

\author[0009-0008-4009-3391]{Varun Bajaj}
\affiliation{Space Telescope Science Institute, 3700 San Martin Drive Baltimore, MD 21218, USA}

\author[0000-0002-2918-7417]{Jenna E. Ryon}
\affiliation{Space Telescope Science Institute, 3700 San Martin Drive Baltimore, MD 21218, USA}

\author[0000-0002-1000-6081]{Sean T. Linden}
\affiliation{Department of Astronomy, 
University of Massachusetts, 710 North Pleasant Street, Amherst, MA 01003, USA}
\affiliation{Steward Observatory, University of Arizona, 933 N. Cherry Avenue, Tucson, AZ 85719, USA}

\author[0000-0001-6464-3257]{Matteo Correnti}
\affiliation{INAF Osservatorio Astronomico di Roma, Via Frascati 33, 00078, Monteporzio Catone, Rome, Italy}
\affiliation{ASI-Space Science Data Center, Via del Politecnico, I-00133, Rome, Italy}

\author[0000-0001-6291-6813]{Michele Cignoni}
\affiliation{Department of Physics - University of Pisa, Largo B. Pontecorvo 3, 56127 Pisa, Italy }
\affiliation{INFN, Largo B. Pontecorvo 3, 56127 Pisa, Italy }
\affiliation{INAF - Osservatorio di Astrofisica e Scienza dello Spazio di Bologna, Via Gobetti 93/3, I-40129 Bologna, Italy }

\author[0000-0003-1427-2456]{Matteo Messa}
\affiliation{INAF - Osservatorio di Astrofisica e Scienza dello Spazio di Bologna, Via Gobetti 93/3, I-40129 Bologna, Italy }

\author[0000-0003-2954-7643]{Elena Sabbi}
\affiliation{Space Telescope Science Institute, 3700 San Martin Drive Baltimore, MD 21218, USA}

\author[0000-0001-8608-0408]{John S. Gallagher}
\affiliation{Department of Astronomy, University of Wisconsin-Madison, 475 N. Charter Street, Madison, WI 53706, USA}

\author[0000-0002-3247-5321]{Kathryn Grasha}
\affiliation{Research School of Astronomy and Astrophysics, Australian National University, Canberra, ACT 2611, Australia}   
\affiliation{ARC Centre of Excellence for All Sky Astrophysics in 3 Dimensions (ASTRO 3D), Australia}   
\affiliation{Visiting Fellow, Harvard-Smithsonian Center for Astrophysics, 60 Garden Street, Cambridge, MA 02138, USA}  

\author[0000-0002-8222-8986]{Alex Pedrini}
\affiliation{Department of Astronomy, The Oskar Klein Centre, Stockholm University, AlbaNova, SE-10691 Stockholm, Sweden}

\author[0000-0002-6447-899X]{Robert A. Gutermuth}
\affiliation{Department of Astronomy, 
University of Massachusetts, 710 North Pleasant Street, Amherst, MA 01003, USA}

\author[0000-0003-0470-8754]{Jens Melinder}
\affiliation{Department of Astronomy, The Oskar Klein Centre, Stockholm University, AlbaNova, SE-10691 Stockholm, Sweden}

\author[0000-0002-4460-9892]{Ralf Kotulla}
\affiliation{Department of Astronomy, University of Wisconsin-Madison, 475 N. Charter Street, Madison, WI 53706, USA}

\author[0000-0003-3880-8075]{Gustavo P\'erez}
\affiliation{College of Information and Computer Sciences, University of Massachusetts, 140 Governors Drive, Amherst, MA 01003, USA}

\author[0000-0003-3893-854X]{Mark R. Krumholz}
\affiliation{Research School of Astronomy and Astrophysics, Australian National University, Canberra, ACT 2611, Australia} 

\author[0000-0001-8068-0891]{Arjan Bik}
\affiliation{Department of Astronomy, The Oskar Klein Centre, Stockholm University, AlbaNova, SE-10691 Stockholm, Sweden}

\author[0000-0002-3005-1349]{G\"{o}ran \"{O}stlin}
\affiliation{Department of Astronomy, The Oskar Klein Centre, Stockholm University, AlbaNova, SE-10691 Stockholm, Sweden}

\author[0000-0001-8348-2671]{Kelsey E. Johnson}
\affiliation{Department of Astronomy, University of Virginia, Charlottesville, VA 22904, USA }

\author[0009-0003-6182-8928]{Giacomo Bortolini}
\affiliation{Department of Astronomy, The Oskar Klein Centre, Stockholm University, AlbaNova, SE-10691 Stockholm, Sweden}
\affiliation{Department of Physics - University of Pisa, Largo B. Pontecorvo 3, 56127 Pisa, Italy }

\author[0000-0002-0806-168X]{Linda J. Smith}
\affiliation{Space Telescope Science Institute, 3700 San Martin Drive Baltimore, MD 21218, USA}

\author[0000-0002-0986-4759]{Monica Tosi}
\affiliation{INAF - Osservatorio di Astrofisica e Scienza dello Spazio di Bologna, Via Gobetti 93/3, I-40129 Bologna, Italy }

\author[0000-0002-3869-9334]{Subhransu Maji}
\affiliation{College of Information and Computer Sciences, University of Massachusetts, 140 Governors Drive, Amherst, MA 01003, USA}

\author[0000-0002-2199-0977]{Helena Faustino Vieira}
\affiliation{Cardiff Hub for Astrophysics Research and Technology (CHART), School of Physics \& Astronomy, Cardiff University, The Parade, Cardiff CF24 3AA, UK}

\received{Feb. 29 2024}
\revised{May 3 2024}
\accepted{May 13 2024}
\submitjournal{ApJ}

\begin{abstract}
We present maps of ionized gas (traced by Pa$\alpha$ and Br$\alpha$) and 3.3 $\mu$m Polycyclic Aromatic Hydrocarbon (PAH) emission in the nearby spiral galaxy NGC 628, derived from new JWST/NIRCam data from the FEAST survey. With this data, we investigate and calibrate the relation between 3.3 $\mu$m PAH emission and star formation rate (SFR) in and around emerging young star clusters (eYSCs) on a scale of ${\sim}40$ pc. We find a tight (correlation coefficient $\rho$${\sim}$0.9) sub-linear (power-law exponent $\alpha$${\sim}$0.75) relation between the 3.3 $\mu$m PAH luminosity surface density and SFR traced by Br$\alpha$ for compact, cospatial (within 0.16$''$ or ${\sim}$7 pc) peaks in Pa$\alpha$, Br$\alpha$, and 3.3 $\mu$m (eYSC--I). The scatter in the relationship does not correlate well with variations in local interstellar medium (ISM) metallicity due to a radial metallicity gradient, but rather is likely due to stochastic sampling of the stellar initial mass function (IMF) and variations in the PAH heating and age of our sources. The deviation from a linear relation may be explained by PAH destruction in more intense ionizing environments, variations in age, and IMF stochasticity at intermediate to low luminosities. We test our results with various continuum subtraction techniques using combinations of NIRCam bands and find that they remain robust with only minor differences in the derived slope and intercept. An unexpected discrepancy is identified between the relations of hydrogen recombination lines (Pa$\alpha$ versus Br$\alpha$; H$\alpha$ versus Br$\alpha$).
\end{abstract} 


\hypertarget{1}{\section{Introduction}}

The calibration of short wavelength ($\lesssim 8 \, \mu$m) tracers of dust-obscured star formation (SF) is becoming increasingly important in the era of the James Webb Space Telescope (JWST). Standard mid-infrared (MIR) star formation rate (SFR) indicators, e.g. the bright Polycyclic Aromatic Hydrocarbon (PAH) emission around 8 $\mu$m, are observable with JWST/MIRI only out to a redshift z$\sim$2. Yet, it is well established that more than half of the SFR budget in galaxies at z$\leq$5 is obscured by dust and emerges in the infrared (IR) \citep{2018ApJ...862...77C,2020ApJ...902..112B}, highlighting the importance of accounting for the dust-obscured SFR component at low and intermediate redshifts. The 3.3 $\mu$m PAH emission feature is a strong candidate to push dust-obscured SFR estimates beyond z${\sim}2$ and out to z${\sim}7$ with JWST/MIRI \citep[e.g.][]{2020ApJ...905...55L}.

PAHs are small dust grains associated with the ubiquitously observed near-infrared (NIR) and MIR emission features in galaxies \citep[e.g.][]{1984A&A...137L...5L,1985ApJ...290L..25A,1989ApJS...71..733A,2008ARA&A..46..289T}. These NIR/MIR features are the result of the de-excitation of PAH grains through vibrational modes of C-H and C-C bonds after the absorption of an ultraviolet (UV) or optical photon, typically in the energy range of ${\sim}$3-9 eV \citep[see][]{2021ApJ...917....3D}. These emission features are very bright in typical galaxies, consisting of $10-20\%$ of the total IR emission \citep{2000ApJ...532L..21H,2007ApJ...656..770S,2008ARA&A..46..289T,2020NatAs...4..339L}. 

PAH grains are fragile and are destroyed by the ionizing radiation from newly formed stars, requiring shielding from larger grains to survive \citep{2004ApJS..154..253H,2007ApJ...660..346P,2008MNRAS.389..629B,2009ApJ...699.1125R}. For this reason, PAHs are heated and emit in the photodissociation regions (PDRs) that surround star-forming regions, but not inside them \citep[e.g.][]{2009ApJ...699.1125R}. Due to the tight spatial connection between PAH emission and active star-forming regions in galaxies, PAHs have been widely used as SFR tracers in the past \citep[e.g.][]{2004ApJS..154..253H,2004ApJ...613..986P,2006ApJ...652..283B,2007ApJ...666..870C,2007ApJ...657..810D,2007ApJ...656..770S,2009ApJ...703.1672K,2016ApJ...818...60S}. In particular, the brightest PAH emission feature at about 8 (or 7.7) $\mu$m has been used out to high redshift \citep[e.g.][]{2011A&A...533A.119E}. Recently with data from JWST/MIRI, Hubble Space Telescope (HST), Spitzer, and Herschel, \cite{2023arXiv231007766R} study the relationship between 8 $\mu$m PAH luminosity and UV-derived SFR in a sample of galaxies at z${\sim}0\,$-$\,2$ using spectral energy distribution (SED) modeling and find a tight correlation between the rest-frame dust-corrected far-UV (FUV) and MIRI/F770W luminosities, which they use to calibrate the F770W as a SFR tracer. 

However, the abundance and emission of PAHs can depend on galaxy properties such as metallicity and star formation history, which complicates their use as SFR indicators. Numerous studies have observed a deficit in the PAH luminosity at 8 $\mu$m in low metallicity galaxies  \citep[e.g.][]{2005ApJ...628L..29E,2007ApJ...666..870C,2007ApJ...656..770S,2014MNRAS.445..899C,2017ApJ...837..157S,2022ApJ...928..120G}, where there are less metals available in the interstellar medium (ISM) to shield PAH grains. \cite{2012ApJ...744...20S} argue that metals act as catalysts for the formation and growth of PAHs, which leads to smaller average PAH sizes in metal-poor environments and thus a higher probability of destruction under normal ISM conditions. \cite{2022ApJ...928..120G} find that on scales of ${\sim}$1 kpc, the 8 $\mu$m luminosity varies by over a factor of 10 at low surface densities of SFR and $>\,$3 at high SFR, and that the variation correlates well with differences in metallicity between galaxies. In addition, the existence of a strong interstellar radiation field is found to suppress PAH emission \citep{2006A&A...446..877M,2008ApJ...682..336G,2011ApJ...728...45L,2017ApJ...837..157S,2018ApJ...864..136B}. Further complicating the picture, spatially resolved studies show that a fraction of the 8 $\mu$m emission is associated with the diffuse ISM, suggesting an additional heating component other than recent ($<100$ Myr) star formation \citep{2008MNRAS.389..629B,2014ApJ...784..130C,2014ApJ...797..129L}.

The 3.3 $\mu$m PAH emission in particular originates from the radiative relaxation of C-H stretching modes of small, neutral PAHs \citep{2004ApJ...611..928V,2020MNRAS.494..642M} and represents ${\sim}0.1\%$ of the total IR power and ${\sim}1.5\%-3\%$ of the total PAH emission in galaxies \citep{2020ApJ...905...55L}. The feature has been historically very difficult to observe in galaxies. The Spitzer IRAC 3.6 $\mu$m band contains the 3.3 $\mu$m PAH, but is dominated by stellar emission in all but the densest of regions \citep[e.g.][]{2015ApJS..219....5Q}. Spectroscopic surveys of the feature in external galaxies by the Infrared Space Observatory (ISO) and AKARI have been limited to the brightest sources. As a result, the 3.3 $\mu$m PAH feature has yet to be properly calibrated as a SFR tracer, except in a sample of nearby PAH-bright galaxies, mostly consisting of luminous/ultraluminous infrared galaxies (LIRGs/ULIRGs) \citep{2020ApJ...905...55L}. The calibration of the 3.3 $\mu$m PAH feature as a SFR indicator in typical star-forming galaxies requires the sensitivity and imaging capabilities of JWST and its more targeted filter selection (e.g. the NIRCam/F335M). 

The relative strength of the 3.3 $\mu$m PAH emission as a tracer of SFR comes from the fact that it is about 2.5 times less sensitive to dust extinction than Pa$\alpha$ (1.87 $\mu$m) and 3-10 times brighter than Br$\alpha$ (4.05 $\mu$m) in typical galaxies \citep{2018A&A...617A.130I}, making it easier to detect at high redshift. This emission feature has been detected at high redshift with Spitzer MIR spectroscopy of ULIRGs at z${\sim}$2 \citep{2009ApJ...703..270S} and in a strongly lensed galaxy at z${\sim}$3 \citep{2009ApJ...698.1273S}. With JWST/MIRI spectroscopy, the 3.3 $\mu$m PAH feature can be observed out to z${\sim}7$ before being shifted out of the wavelength coverage. However, due to the much lower sensitivity of MIRI/MRS channel 4, a more reasonable expectation of the highest detectable redshift of 3.3 $\mu$m with JWST is z${\sim}4.5$, corresponding to the longest wavelength of channel 3. \cite{2023Natur.618..708S} report the detection of the 3.3 $\mu$m PAH feature in a z${\sim}4.2$ galaxy using MIRI/MRS spectroscopy, making the most distant detection of PAH emission to date.

Observationally, little is currently known about the nature of the 3.3 $\mu$m PAH emission. Results from one metal-poor (0.25 Z$_{\odot}$) galaxy indicate that the 3.3 $\mu$m PAH emission is relatively stronger than the emission from other PAH features in low metallicity (or higher ionizing) environments, suggesting (1) a shift in the size distribution towards smaller PAHs possibly due to the better survivability of small grains via efficient relaxation by recurrent fluorescence \citep[see][]{1988PhRvL..60..921L,2017MNRAS.469.4933L,2020Ap&SS.365...58W}, or (2) the shattering of large PAH grains into smaller ones, or (3) that the emission from large PAH grains is shifted to shorter wavelengths in intense environments \citep{2020ApJ...905...55L}. New studies from JWST are beginning to shed light on the 3.3 $\mu$m PAH feature. \cite{2023ApJ...957L..26L} study the 3.3 $\mu$m PAH emission in a LIRG at a distance of ${\sim}70$ Mpc with JWST/NIRSpec/IFU and find suppression in the PAH emission relative to the ionized gas in the central 1 kpc region of the active galactic nucleus (AGN) and clear differences in average grain properties, suggesting smaller grains are preferentially destroyed in the vicinity of the AGN. \cite{2023ApJ...944L...7S} use the JWST/NIRCam medium bands F300M, F335M and F360M to derive maps of the 3.3 $\mu$m PAH emission across three nearby galaxies and find that the PAH-to-continuum ratios for F335M are between 5\% and 65\% and increase smoothly with galactocentric radius outside of the galaxy centers. \cite{2023ApJ...944L..12C} utilize these 3.3 $\mu$m PAH maps in combination with the features traced by MIRI F770W and F1130W and find based on the PAH ratios that HII regions/more ionized environments may be populated by hotter or smaller PAHs, and have larger PAH ionization fractions.

In addition to the emission from PAHs, ionized gas emission is used to estimate SFRs in galaxies. Young ($<$10 Myr), massive ($>$15 $M_{\odot}$) stars produce an abundance of high-energy photons that ionize the surrounding gas.  As this ionized gas cools, hydrogen recombines and emits a series of emission lines as the electron settles to the ground state. Hydrogen recombination lines such as H$\alpha$ and H$\beta$ are strong in star-forming galaxies and have been used extensively in the past to study star formation by tracing the ionizing photon rate \citep{1998ARA&A..36..189K}. Yet, in dense star-forming regions, recombination lines in the optical regime like H$\alpha$ and H$\beta$ can be significantly affected by extinction. The ratios of various Balmer series lines (known as the Balmer decrement) can be utilized to correct the effects of dust. This has been applied in a number of large surveys of nearby galaxies \citep[e.g.][]{2002AJ....124.3135K,2004MNRAS.351.1151B,2006ApJ...642..775M}. This approach is generally only effective at low to moderate extinction and has been shown to underestimate the obscured SFR \citep[e.g.][]{2022ApJS..263...17G}. Alternatively, longer wavelength recombination lines can be used, such as Pa$\alpha$, Pa$\beta$, and Br$\alpha$. These lines are emitted at NIR/MIR wavelengths and as a result, suffer from significantly less dust obscuration, but are many times fainter. Long wavelength recombination lines are also more sensitive to the physical conditions of the gas. For Br$\alpha$, the variations are ${\sim}58\%$ for an electron temperature (T$_{e}$) in the range 5000--20000 K, and ${\sim}13\%$ for density ($n_{e}$) in the range $10^{2}$--$10^{6}$ cm$^{-3}$ \citep{2013seg..book..419C}.

In this study, we map the ionized gas (Pa$\alpha$ and Br$\alpha$) and PAH (3.3 $\mu$m) emission across the galaxy NGC 628 at the angular resolution of ${\sim}{\,}$0.07--0.15$''$ (${\sim}$3--7 pc) using new JWST/NIRCam data from the FEAST (Feedback in Emerging extrAgalactic Star clusTers) survey. Catalogs of candidate young and embedded star clusters are extracted as peaks in both the ionized gas and PAH maps. For these sources, we measure the PAH and ionized gas luminosities and evaluate the relationship between them. Our goal is to provide an initial calibration of the relation between 3.3 $\mu$m PAH emission and SFR in this one system, but later refine this with more data that will provide the necessary handle on the expected sources of variation/uncertainty. 

The galaxy NGC 628 (M74) is a well-studied grand-design spiral that lacks a central bar. It is located at a distance of 9.84 Mpc \citep{2009AJ....138..332J,2021MNRAS.501.3621A}, with a near face-on orientation \citep[inclination $i=8.9^{\circ}$;][]{2020ApJ...897..122L}. NGC 628 is actively star-forming with a mean SFR surface density of 0.003 M$_{\odot}$ yr$^{-1}$ kpc$^{-2}$ and a total SFR of 0.7  M$_{\odot}$ yr$^{-1}$ \citep{2010ApJ...714.1256C}. The stellar mass is log(M$_{*}$/M$_{\odot}$) = 10.34 \citep{2021ApJS..257...43L}, placing NGC 628 as a typical main-sequence galaxy in the local volume \citep[e.g.][]{2014MNRAS.445..899C}. Its characteristic metallicity, defined as the oxygen abundance at 0.4$\,$R$_{25}$ (where R$_{25}$ is the radius in the B band equal to 25 mag), is 12+log(O/H) = 8.55 \citep{2020ApJ...893...96B}. \cite{2020ApJ...893...96B} find a moderate radial metallicity gradient of $-0.4$ dex R$_{25}^{-1}$ or $-0.03$ dex kpc$^{-1}$ at the distance listed above and R$_{25}$=315$''$ \citep{2011MNRAS.414..538K}. NGC 628 is also rich in molecular gas with a total molecular gas mass log(M$_{mol}$)= 9.47 M$_{\odot}$ \citep{2021ApJS..257...43L}, assuming $\alpha_{CO}$=4.35 M$_{\odot}$ pc$^{-2}$ (K km s$^{-1}$)$^{-1}$ \citep{2013Natur.499..450B} and a CO(2--1)-to-CO(1--0) ratio of R$_{21}$ = 0.65 \citep{2013AJ....146...19L}. Recent JWST/MIRI observations of the PAH emission across NGC 628 have uncovered an intricate network of filaments of dust emission that tightly trace molecular gas structure \citep{2023ApJ...944L...9L} and dust attenuation \citep{2023ApJ...944L..13T}. These new observations also unveil that the ISM of NGC 628 is characterized by giant bubbles driven by stellar feedback \citep{2023ApJ...944L..22B,2023ApJ...944L..24W}.

This paper is organized as follows. In Section \hyperlink{2}{2}, the JWST NIRCam data is presented, along with the basic data reduction. In Section \hyperlink{3}{3}, the analysis of our data is described, including continuum subtraction techniques, selecting emerging star clusters, and measuring aperture photometry. We present the results of our work in Section \hyperlink{4}{4}, including a novel calibration of the 3.3 $\mu$m PAH emission as a SFR indicator. In Section \hyperlink{5}{5}, we discuss our results and their implications in the context of previous work. In Section \hyperlink{6}{6}, we highlight our main conclusions. In Appendix \hyperlink{A}{A}, we test a variety of continuum subtraction techniques and how they affect our results. In Appendix \hyperlink{B}{B}, we present the results of a binning analysis of our sources.

\hypertarget{2}{\section{Data}}

The JWST data used in this study were obtained in Cycle 1 as part of the JWST--FEAST program (ID 1783, PI: A. Adamo). For the first galaxy observed of the five targets in the sample, NGC 628, we utilize JWST/NIRCam imaging with various filters including F150W, F187N, F200W, F277W, F335M, F405N, and F444W. We obtain stage two calibrated data products from the Mikulski Archive for Space Telescopes (MAST), produced via the NIRCam calibration pipeline 1.12.5 using the calibration reference data context number 1169. Catalogs containing point spread function (PSF) fit positions and fluxes are extracted from the stage two products using the Python package \texttt{one$\_$pass$\_$fitting}\footnote{\url{https://github.com/Vb2341/One-Pass-Fitting}}, with PSF models created by WebbPSF \citep{2014SPIE.9143E..3XP}.  Due to the difference in wavelength and sensitivity, the NIRCam images and corresponding catalogs cannot be consistently aligned to Gaia \citep{2016A&A...595A...1G,2023A&A...674A...1G}.  Therefore, we use a laddered approach. We first align the archival HST/ACS F814W image, obtained by programs 9796 (PI: Miller, J.) and 10402 (PI: Chandar, R.), to Gaia \citep[see][]{2017wfc..rept...19B} and then extract a catalog from F814W on the Gaia astrometric frame.  We provide this reference F814W catalog and the PSF fit NIRCam F200W catalogs as custom, user-supplied catalogs in stage three of the JWST calibration pipeline. The resulting aligned F200W catalogs are combined and used as a reference for the remaining NIRCam data, yielding an overall astrometric precision of 10 milliarcseconds or less. Aligned NIRCam images are then combined into a single mosaic for each filter using the stage three pipeline, projected onto the same pixel grid with a scale of 0.04$''$/pixel.  The resulting mosaics are converted from units of MJy/sr to Jy/pixel. A more detailed description of the data reduction process is presented in Adamo et al. (in prep.). 

\begin{figure}
\centering
\includegraphics[width=0.47\textwidth]{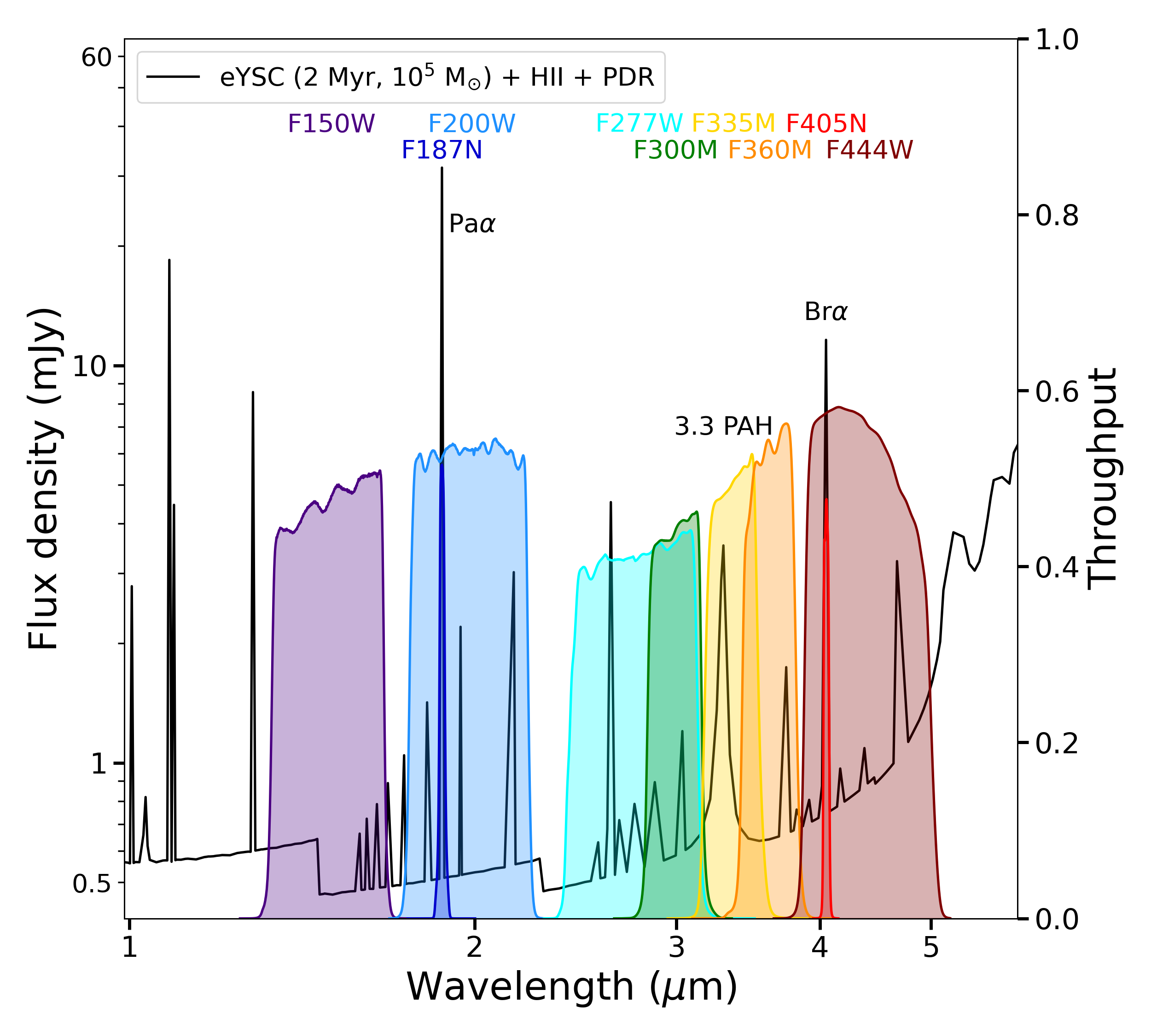}
\caption{A model spectrum of a 2 Myr old, 10$^{5}$ M$_{\odot}$ eYSC powering both an HII region and PDR from the MAPPINGS III derived models of \cite{2008ApJS..176..438G} (black line). Overlaid on top are the NIRCam filter throughputs (colored curves), outlining the regions of the spectrum sampled by our data. The F187N and F405N target the Pa$\alpha$ and Br$\alpha$ hydrogen recombination lines, respectively. The F335M targets the 3.3 $\mu$m PAH emission feature. All other filters target the continuum emission. }
\label{fig:spec}
\end{figure}

Figure \ref{fig:spec} shows the average total system throughput curves for all NIRCam filters investigated in this study, on top of a representative model spectrum, corresponding to a 2 Myr old, 10$^{5}$ M$_{\odot}$ eYSC generating both an HII region and the surrounding PDR from the work of \cite{2008ApJS..176..438G}. This model assumes Starburst99 models \citep{1999ApJS..123....3L} as input stellar spectra, a one-dimensional dynamical evolution model of HII regions, and the MAPPINGS III photoionization code \citep{2004PhDT.......183G} to generate the SEDs. The model spectrum may roughly represent the sources investigated in this study and thus provides an outline of the regions of the spectrum sampled by the NIRCam filters.

\begin{figure*}
\centering
\includegraphics[width=0.95\textwidth]{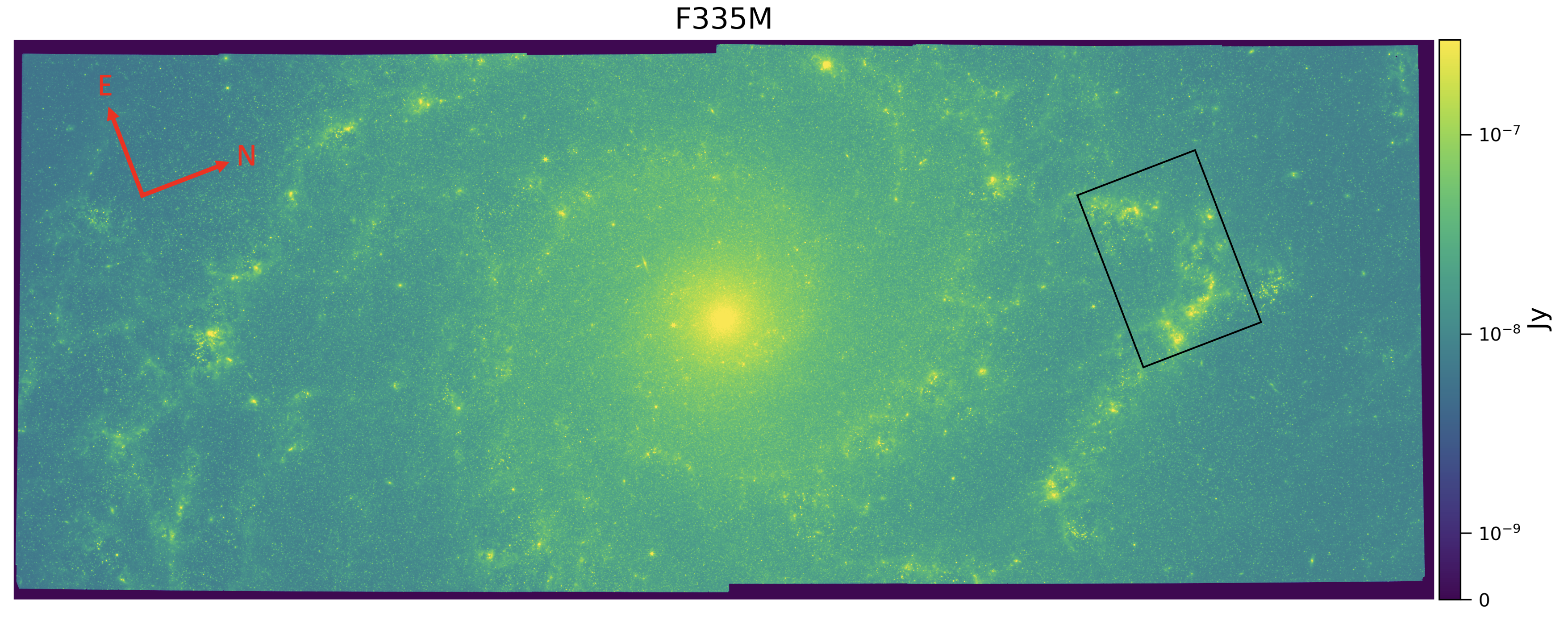}
\includegraphics[width=0.97\textwidth]{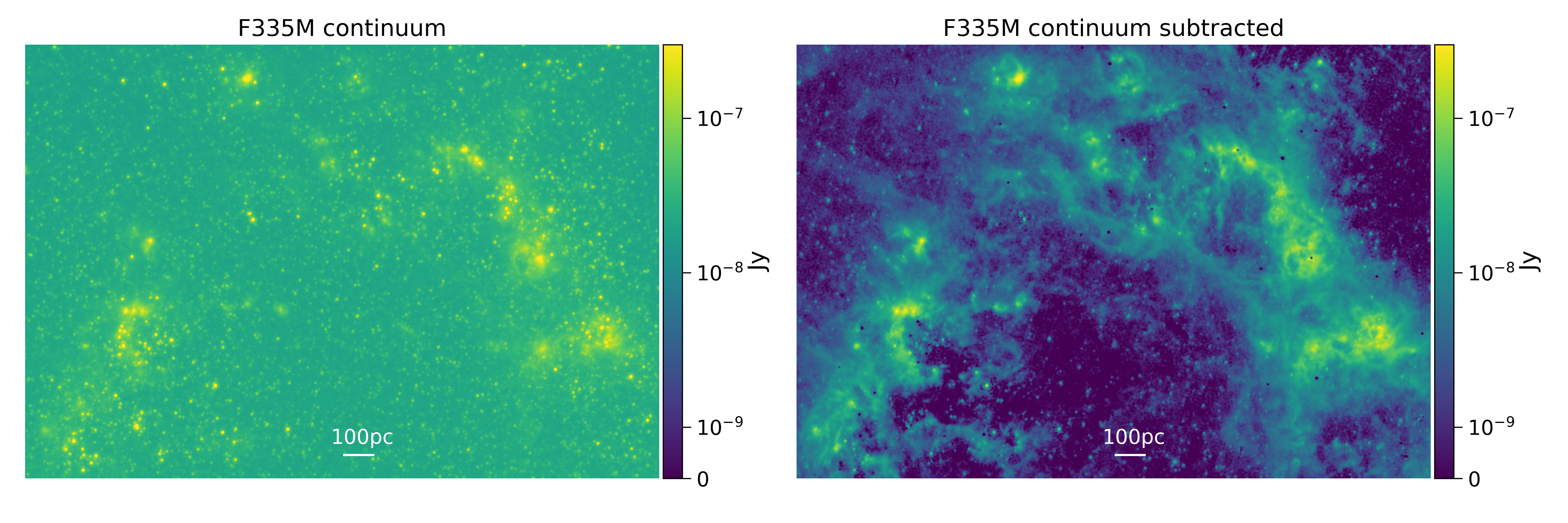}
\caption{ Top panel: the reduced JWST/NIRCam F335M image of NGC 628 ($5.9\arcmin\times2.3\arcmin$ or 16.8 kpc$\times6.5$ kpc) at full resolution. Bottom panels: zoom-ins (rotated; North up, East left) for a representative region shown by the black rectangle in the top panel. Left: the continuum in the F335M filter derived by the method presented in Section \hyperlink{3.1}{3.1}. This includes interpolating the SED between the F277W and F444W filters and scaling by a factor of 1.06. Each image is matched to the PSF of F444W. The contribution of Br$\alpha$ to the F444W filter has been removed. Right: the continuum subtracted F335M image. The white bar shows the scale corresponding to 100 pc for our adopted distance. }
\label{fig:f1}
\end{figure*}

The top panel of Figure \ref{fig:f1} shows the reduced NIRCam F335M image of NGC 628 at native resolution, corresponding to a field of view (FOV) of $5.9\arcmin\times2.3\arcmin$ or 16.8 kpc$\times6.5$ kpc. The F335M filter is centered on the 3.3 $\mu$m PAH emission feature (see Figure \ref{fig:spec}) but also receives contributions from the continuum emission of low mass and/or evolved stars and the continuum from both hot dust grains heated in dense star-forming regions by newly formed stars and a more diffuse dust that is stochastically heated by any stellar population. In star-forming regions, the F335M is expected to be dominated by the bright 3.3 $\mu$m PAH emission feature, but the sources of the continuum emission must be properly accounted for.

Two other complementary NIRCam bands have been observed for NGC 628 and are publicly available on MAST, obtained as part of the PHANGS--JWST program (ID 2107; PI: J. C. Lee) and presented in \cite{2023ApJ...944L..17L}. We acquire the stage two NIRCam F300M and F360M data products from this program and run them through our data reduction process. These observations overlap almost completely with the larger footprint of the FEAST NIRCam mosaics, but only cover a fraction of the field of view ($4.3\arcmin\times2.2\arcmin$ or 12.3 kpc$\times6.3$ kpc). With these additional bands, we are able to explore a variety of different continuum subtraction recipes for the 3.3 $\mu$m PAH emission. Additionally, we align the archival HST/ACS F555W and F658N images of NGC 628, obtained by HST programs 9796 (PI: Miller, J.) and 10402 (PI: Chandar, R.), to the same F814W image used as a reference for NIRCam. These HST images are matched to the same astrometric frame and sampling as the NIRCam data and are used to trace the H$\alpha$ emission across the majority of our mosaics.

For this study, we match the PSFs of each band to the common, lowest-resolution PSF of NIRCam/F444W, which has a full-width-at-half-maximum (FWHM) of 0.145$''$ \citep{2023PASP..135d8001R}. This is important for the accurate subtraction of point sources (e.g. stars) in the derived emission line maps and to allow for a direct comparison between the various emission lines. To do this, we use the effective PSFs \citep[ePSFs; see][]{2000PASP..112.1360A} described in detail in Adamo et al. (in prep.). These are based on the PSF models of \cite{2006acs..rept....1A}\footnote{\url{https://www.stsci.edu/~jayander/HST1PASS/LIB/PSFs/STDPSFs/}} for HST/ACS and WebbPSF \citep{2014SPIE.9143E..3XP} for JWST. Grids of these PSF models are placed in blank copies of the individual frames for each filter, which are then drizzled together using the same parameters as our science images. PSFs are extracted from these drizzled frames and are combined to create the ePSF for each filter. From these ePSFs, we create convolution kernels for each filter to the PSF of F444W via the method described in \cite{2011PASP..123.1218A} as implemented in the \texttt{make$\_$jwst$\_$kernels}\footnote{\url{https://github.com/thomaswilliamsastro/jwst_beam_matching}} code. The reduced science images are convolved with these kernels to create a dataset with all filters matched to the PSF of F444W. The full-resolution, non-convolved images are also used in this study, but only for the selection and cleaning of source catalogs from the emission line maps.

\hypertarget{3}{\section{Analysis}}
\hypertarget{3.1}{\subsection{Continuum subtractions}}

From the fully reduced and processed JWST/NIRCam and HST/ACS images, we create continuum subtracted emission line maps for the hydrogen recombination lines Pa$\alpha$, Br$\alpha$, and H$\alpha$ and the 3.3 $\mu$m PAH emission feature. Our subtraction method utilizes a shorter and a longer wavelength filter to derive the continuum in the emission line filter. For instance, we estimate the continuum in the F335M filter at each pixel in the image by linearly interpolating the SED between the F277W and F444W filters at the location of F335M. The resulting continuum image is then subtracted from the F335M to derive the 3.3 $\mu$m PAH emission line map. To produce the Pa$\alpha$ emission line map, we use the F150W and F200W filters to remove the continuum at F187N. For Br$\alpha$, the F277W and F444W filters are used to remove the continuum at F405N. See Figure \ref{fig:spec} for a visual representation of the emission probed by each NIRCam filter. For H$\alpha$, we use the HST/ACS F555W and F814W filters to remove the F658N continuum.

There are a few complications in our method of isolating the emission lines from the underlying continuum. For one, the F200W filter is contaminated by the Pa$\alpha$ emission line, while the F444W is contaminated by Br$\alpha$ (see Figure \ref{fig:spec}). We implement an iterative subtraction technique to remove the contribution of these lines to the continuum tracing filters. For instance, we derive a F200W image that is corrected for the contribution from Pa$\alpha$ by scaling the initial continuum subtracted F187N image by the ratio of the bandwidths between F187N and F200W and then subtracting this from the F200W. We then use this corrected F200W image to perform an updated continuum subtraction of F187N. The process is repeated until the mean relative difference between the iterations in star-forming regions is less than $10^{-4}$, which is achieved in 3 iterations. We find the correction to the F200W and F444W filters to be relatively small, ${\sim}3\%$ on average in star-forming regions.

The F335M band is centered on the strong aromatic 3.3 $\mu$m feature, but also includes a contribution from the much weaker \citep[${\sim}$0.1$-$0.2 times as bright in the presence of star formation;][]{2012A&A...541A..10Y} aliphatic 3.4 $\mu$m feature and 3.47 $\mu$m plateau feature. Our final 3.3 $\mu$m PAH emission line maps receive a contribution from these other emission features, but we expect the maps to be dominated by the aromatic 3.3 $\mu$m feature within star-forming regions. Separating out these additional components of the F335M and determining their relative contribution within the embedded, young sources investigated in this study will require spectroscopy, which is possible with JWST/NIRSpec. 

For the 3.3 $\mu$m PAH emission line maps, we achieve a better subtraction of the stars when the continuum image is scaled up prior to the subtraction, while for Br$\alpha$, we find that no scaling is needed. This can likely be explained by the fact that F335M is located near the center of the wavelength range between the two continuum tracing filters, F277W and F444W. In this case, the effect due to nonlinearity in the SED across this wavelength range will be maximized, leading to a mismatch between the estimated and true continuum that can be accounted for by introducing a scaling factor. On the other hand, the F405N is very near in wavelength to F444W and the effect due to nonlinearity will be minimal. We expect the F277W to be mostly dominated by stellar continuum, while in regions of star formation, the F444W may be dominated by the continuum from hot dust. For regions in the galaxy with a large stellar contribution, such as in the central bulge, the interpolation between F277W and F444W can overestimate the continuum in the F335M and F405N filters. This can cause a slight oversubtraction of the continuum across these regions in the derived emission line maps, especially when the continuum image is scaled up before subtraction. The effect will be the largest for F335M. 

We carefully visually inspect a range of scaling factors for the continuum between 1.0 and 1.2 for F335M and determine that scaling the continuum up by a factor of 1.06 before the subtraction strikes the balance between the optimal subtraction of the stars in the field, while also limiting oversubtraction of the continuum in the central regions. This scale factor of 1.06 for the continuum is assumed for the final 3.3 $\mu$m PAH emission line maps used in the rest of this paper. 

In Appendix \hyperlink{A}{A}, we discuss a variety of different continuum subtraction techniques and how they affect our results, including the effect of the assumed continuum scaling factor for the 3.3 $\mu$m PAH emission. In addition, we discuss the use of F300M, expected to be mostly dust continuum emission in regions of star formation \citep[e.g.][]{2021ApJ...917....3D}, in place of F277W for the continuum subtraction of F335M and F405N. We also directly compare our results with the continuum subtraction method for F335M developed by the study of \cite{2023ApJ...944L...7S}, which instead utilizes F300M and F360M as the continuum tracers. The main issue with using F360M as a continuum tracer is that it is contaminated by the 3.4 $\mu$m aliphatic and 3.47 $\mu$m plateau features. The method of \cite{2023ApJ...944L...7S} provides a first-order correction for this contamination by utilizing the observed F335M/F300M and F360M/F300M colors. Our method bypasses these issues by using F444W as the long wavelength continuum tracing filter, however, it may introduce other uncertainties given the longer wavelength of F444W and the rapidly increasing contribution from the dust continuum. Although we observe some minor differences in our results, our major findings remain unaffected by the assumed continuum subtraction method (Appendix \hyperlink{A}{A}). 

\begin{figure*}
\centering
\includegraphics[width=0.9\textwidth]{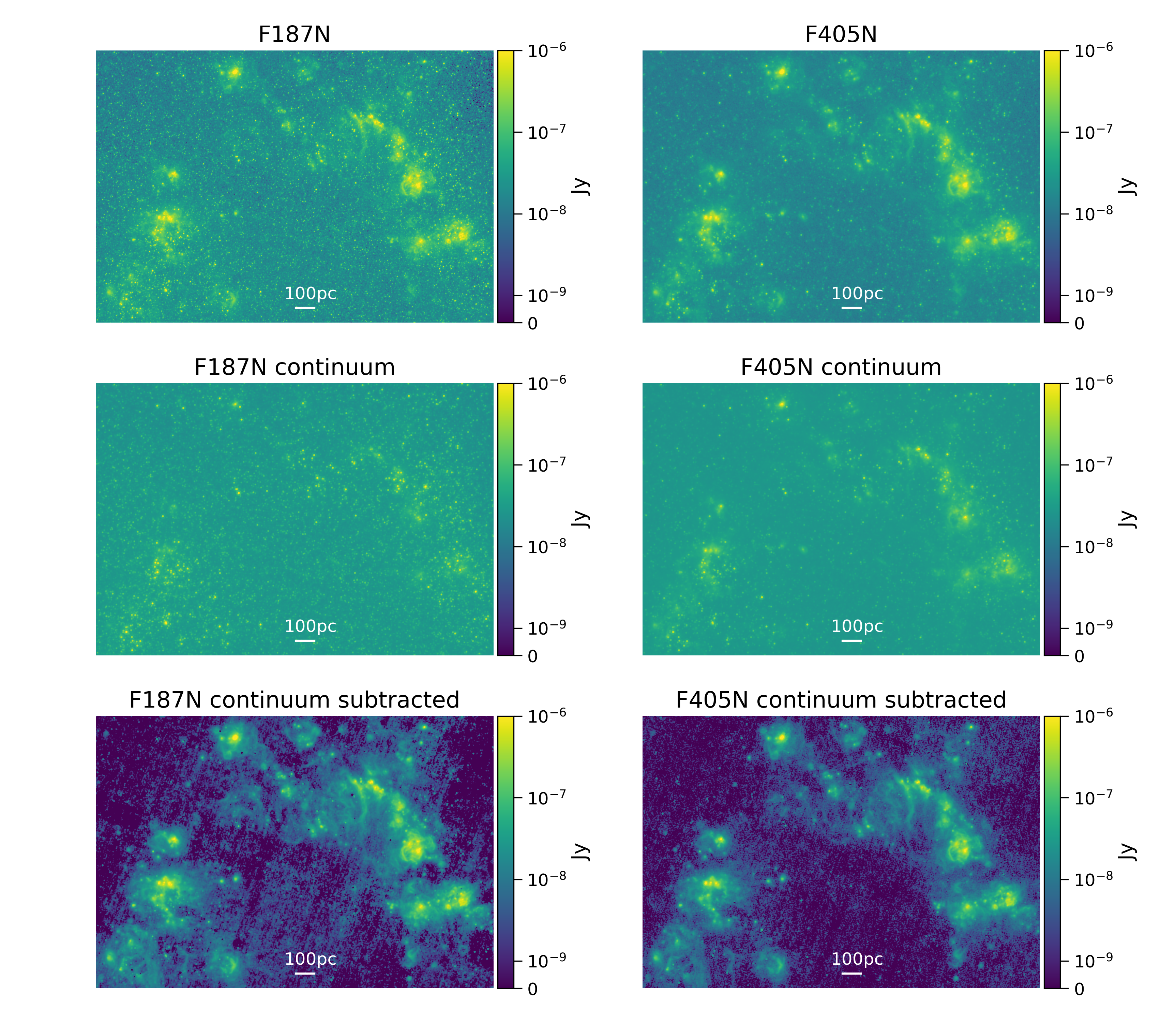}
\caption{ The continuum subtractions of the F187N (Pa$\alpha$; left panels) and the F405N (Br$\alpha$; right panels) for the same representative region as displayed in Figure \ref{fig:f1}. Top panels: the reduced F187N (left) and F405N (right) images at full resolution. Middle panels: the derived continuum in F187N (left) and F405N (right). For F187N, the continuum is interpolated from the F150W and F200W filters, while for F405N it is interpolated from F277W and F444W. Each image is matched to the common/broadest PSF of F444W. The contributions of Pa$\alpha$ to F200W and Br$\alpha$ to F444W have been removed. Bottom panels: the continuum subtracted F187N (left) and F405N (right). The white bar shows the scale corresponding to 100 pc.}
\label{fig:f2}
\end{figure*}

The bottom panels of Figure \ref{fig:f1} show the F335M continuum image (left) and the continuum subtracted F335M (right) for a representative region showing a segment of spiral arms that contains multiple large star-forming regions. These images are derived by the method presented above, i.e. interpolating between F277W and F444W and scaling by a factor of 1.06. The left panels of Figure \ref{fig:f2} show images of the same region for the reduced F187N (top), the F187N continuum (middle), and the continuum subtracted F187N (bottom; Pa$\alpha$). The right panels of Figure \ref{fig:f2} show the same, but for F405N (i.e. Br$\alpha$). These images show that our methods work generally well to remove the stellar and dust continuum from under the emission lines. It is interesting to note here that we see a clear signature of diffuse emission in star-forming regions in F444W, likely originating from the continuum emission of hot dust. As a result, the lower left panel of Figure \ref{fig:f1} (F335M continuum) and the middle right panel of Figure \ref{fig:f2} (F405N continuum) show diffuse emission, attributed to hot dust continuum, while the middle left panel of Figure \ref{fig:f2} (F187N continuum) does not. For each of the emission line maps, we calculate the image uncertainty by determining the iteratively 3$\sigma$ clipped standard deviation within a blank sky region. The uncertainties are 1.69, 1.29, 4.13, and 0.61 nanojanskys (nJy) for the Pa$\alpha$, Br$\alpha$, H$\alpha$, and 3.3 $\mu$m PAH emission line maps, respectively.

\hypertarget{3.2}{\subsection{Selecting emerging sources}}

The resolution of our JWST/NIRCam maps of ionized gas and 3.3 $\mu$m PAH emission (${\sim}{\,}$0.07--0.15$''$ or 3--7 pc at the adopted distance) is well-matched to the average size of individual young, massive star clusters \citep[${\sim}$3 pc; e.g.][]{2017ApJ...841...92R}. Thus, we expect to effectively resolve individual clusters in our images. In this study, we use the catalogs of candidate emerging Young Star Clusters (eYSCs) created by Adamo et al. (in prep.) from our NIRCam images. These sources are independently selected as bright, compact peaks in the 3.3 $\mu$m PAH, Pa$\alpha$, and Br$\alpha$ emission line maps, using the Python library for Source Extraction and Photometry \citep[\texttt{SEP};][]{1996A&AS..117..393B,2016JOSS....1...58B}. In total, the extraction results in 16,217 peaks identified across the three emission line maps. These catalogs are then cleaned visually to remove contaminants such as obvious point sources (e.g. stars), sources on the edge of the mosaics, hot pixels, etc. The extraction and visual inspection are performed on the non-convolved continuum subtracted images as we find it easier to identify and remove contaminants. 

Aperture photometry is measured for these emission line peaks in all the NIRCam images for 4 pixel radius circular apertures, as detailed in Adamo et al. (in prep.). The measurements are corrected for the local background via an annulus with an inner radius of 7 pixels and an outer radius of 9 pixels. In addition, concentration index (CI) based aperture corrections are applied. From these measurements, we limit the catalogs to the bright (magnitude error $\leq$ 0.3), compact peaks. For the Pa$\alpha$ peaks, we require a magnitude error $\leq$ 0.3 in both F187N and F200W. For the Br$\alpha$ and 3.3 $\mu$m PAH peaks, the magnitude error is $\leq$ 0.3 in F405N and F444W, or F335M and F444W, respectively. The error cut on both continuum and emission line filters helps to confirm that the sources are not spurious. ``Compact" refers to the fact that the sources are detected above these thresholds for the 4 pixel (0.16$''$ or ${\sim}$7 pc) radius local background subtracted photometry. We expect these catalogs to give a fairly complete census of the bright, compact ionized gas and PAH emission peaks across our maps. From these catalogs, three distinct classes of sources are selected, defined as eYSC--I, eYSC--II, and PAH compact. Table \hyperlink{t1}{1} gives a summary of the source selection and statistics. 

The eYSC--I sources are selected from the cleaned and cut catalogs as bright Pa$\alpha$ peaks with Br$\alpha$ and 3.3 $\mu$m PAH emission peaks within ($\leq$) 4 pixels or 0.16$''$ (about the FWHM of the F444W PSF). These sources show local, bright peaks in all three emission lines, cospatial within about 7 pc. Based on the tight spatial connection between the compact PAH and ionized gas emission, these sources are expected to be the youngest as they are still embedded in their natal gas and dust. 

The eYSC--II or ``hydrogen recombination line compact" sources are selected as Pa$\alpha$ peaks with a Br$\alpha$ peak within 4 pixels, but no corresponding bright 3.3 $\mu$m PAH emission peak. These sources are expected to be older as they have already mostly emerged from their birth material via feedback mechanisms (e.g. radiation pressure, stellar winds, supernovae). These tend to show more shell-like or filamentary PAH emission in the vicinity. Yet, they are still producing significant amounts of ionizing photons and thus are still young ($<<$ 10 Myr). 

PAH compact sources are selected as 3.3 $\mu$m PAH emission peaks, but with no corresponding bright, compact Pa$\alpha$ or Br$\alpha$ peak within 4 pixels. These sources lack a compact peak in ionized gas emission down to the detection limits, which implies that they do not contain massive stars that ionize the gas. Generally, they are not well detected at optical wavelengths. They may represent older or lower mass clusters. This is due to the fact that PAHs are heated by non-ionizing UV radiation \citep[e.g.][]{2021ApJ...917....3D}, thus not only by massive, young, ionizing sources, but also by lower mass or older, UV-bright sources. These sources are expected to be either too low mass, thus dominated by stochastic sampling of the stellar initial mass function (IMF), or too old to produce significant ionizing photons. The nature of these sources will be further investigated in Linden et al. (in prep.).

\begin{center}
\begin{table}
\centering
\caption{\hypertarget{t1}{ Summary of Source Selection/Statistics}}  
\begin{tabular}{ l | c }
\hline
\hline
\rule{0pt}{3ex}
Total number of sources extracted from \texttt{SEP} & 16217 \\
\rule{0pt}{4ex} Number of Pa$\alpha$ peaks after cleaning  & 1734 \\
\rule{0pt}{2ex} Br$\alpha$ peaks after cleaning  & 1366 \\
\rule{0pt}{2ex} 3.3 $\mu$m PAH emission peaks after cleaning  & 1680 \\
\rule{0pt}{4ex} $^{\mbox{\textit{a}}}$Number of eYSC--I sources & 737 \\
\rule{0pt}{2ex} $^{\mbox{\textit{b}}}$eYSC--II sources & 333 \\[1mm]
\rule{0pt}{2ex} $^{\mbox{\textit{c}}}$PAH compact sources & 638 \\
\rule{0pt}{4ex} Number of eYSC--II isolated by  $\geq$20 pixels & 71 \\
\rule{0pt}{2ex} PAH compact sources isolated by $\geq$20 pixels & 257  \\[1mm]
\hline
\end{tabular}
\begin{flushleft} 
\rule{0pt}{3ex}
\currtabletypesize{\sc Note}--- \\
\rule{0pt}{4ex}
$^{\mbox{\textit{a}}}$ Selected as Pa$\alpha$ peaks, with Br$\alpha$ and 3.3 $\mu$m PAH peaks also within 4 pixels (0.16$''$), and a magnitude error $\leq$ 0.3 in F187N, F200W, F335M, F405N, and F444W. \\
$^{\mbox{\textit{b}}}$ Selected as Pa$\alpha$ peaks, with a Br$\alpha$ peak within 4 pixels, but no corresponding bright 3.3 $\mu$m PAH peak, and a magnitude error $\leq$ 0.3 in F187N, F200W, F405N, and F444W. \\
$^{\mbox{\textit{c}}}$ Selected as 3.3 $\mu$m PAH emission peaks, with no corresponding bright Pa$\alpha$ or Br$\alpha$ peak within 4 pixels, and a magnitude error $\leq$ 0.3 in F335M and F444W. \\
\end{flushleft}
\end{table}
\end{center}

\hypertarget{3.3}{\subsection{Photometry}}

Aperture photometry is measured on each of the emission line maps for the sources described in the previous section using the \texttt{photutils} package. We create 200$\times$200 pixel cutouts centered on the location of each source. The photometry is then measured as the sum in circular apertures 10 pixels (0.4$''$ or $\sim$19 pc) in radius at each source location in each of the emission lines. This aperture size is chosen to be around the distance that we expect the local young, ionizing source to dominate the PAH heating in our brightest source. See Section \hyperlink{5.2}{5.2} for a discussion. In short, the Str\"{o}mgren radius of our brightest HII region is ${\sim}$15 pc and the PAH extent is expected to be about that of the HII region \citep{2019ApJ...876...62C}, so the $\sim$19 pc apertures should be sufficiently large to capture the majority of the PAH heating by the local young ionizing source.

We estimate the local background around each source by measuring the iteratively 3$\sigma$ clipped mode within an encompassing annulus of equal area, with an inner radius of 10 pixels and outer radius of 10$\sqrt{2}$ pixels. This is multiplied by the total number of pixels in the aperture to give the total background in the aperture, which is removed from the measurements. The uncertainties in our photometric measurements are derived as the emission line map uncertainties multiplied by the square root of the number of pixels in the aperture. The resulting 3$\sigma$ detection limits for our measurements are 89.9, 68.7, 219.7, and 32.2 nJy for the Pa$\alpha$, Br$\alpha$, H$\alpha$, and 3.3 $\mu$m PAH emission line maps, respectively. We require that our sources are detected above these limits in Pa$\alpha$, Br$\alpha$, and 3.3 $\mu$m PAH emission.  

For our 10 pixel radius apertures, there will be confusion between the measurements of eYSC--II (hydrogen recombination line compact) and PAH compact sources, which are isolated from a bright peak in the other emission line by $>\,$4 pixels. To limit this confusion, we remove all sources in the eYSC--II catalog that have a corresponding bright peak in the 3.3 $\mu$m PAH emission line map within 20 pixels. Similarly, we remove all sources in the PAH compact catalog that have a bright peak in either the Pa$\alpha$ or Br$\alpha$ emission line map within 20 pixels. Thus for our measurements, we expect the remaining sources in these two catalogs to be isolated, in the sense that there is no corresponding bright, compact peak in the other emission line. For eYSC--II and PAH compact sources, we measure 3$\sigma$ upper limits when the measurements have a signal-to-noise ratio (S/N) $<$ 3. The 3$\sigma$ upper limits correspond to the square root of the number of pixels in the aperture multiplied by 3 times the standard deviation within the aperture. 

\begin{figure*}
\centering
\includegraphics[width=0.7\textwidth]{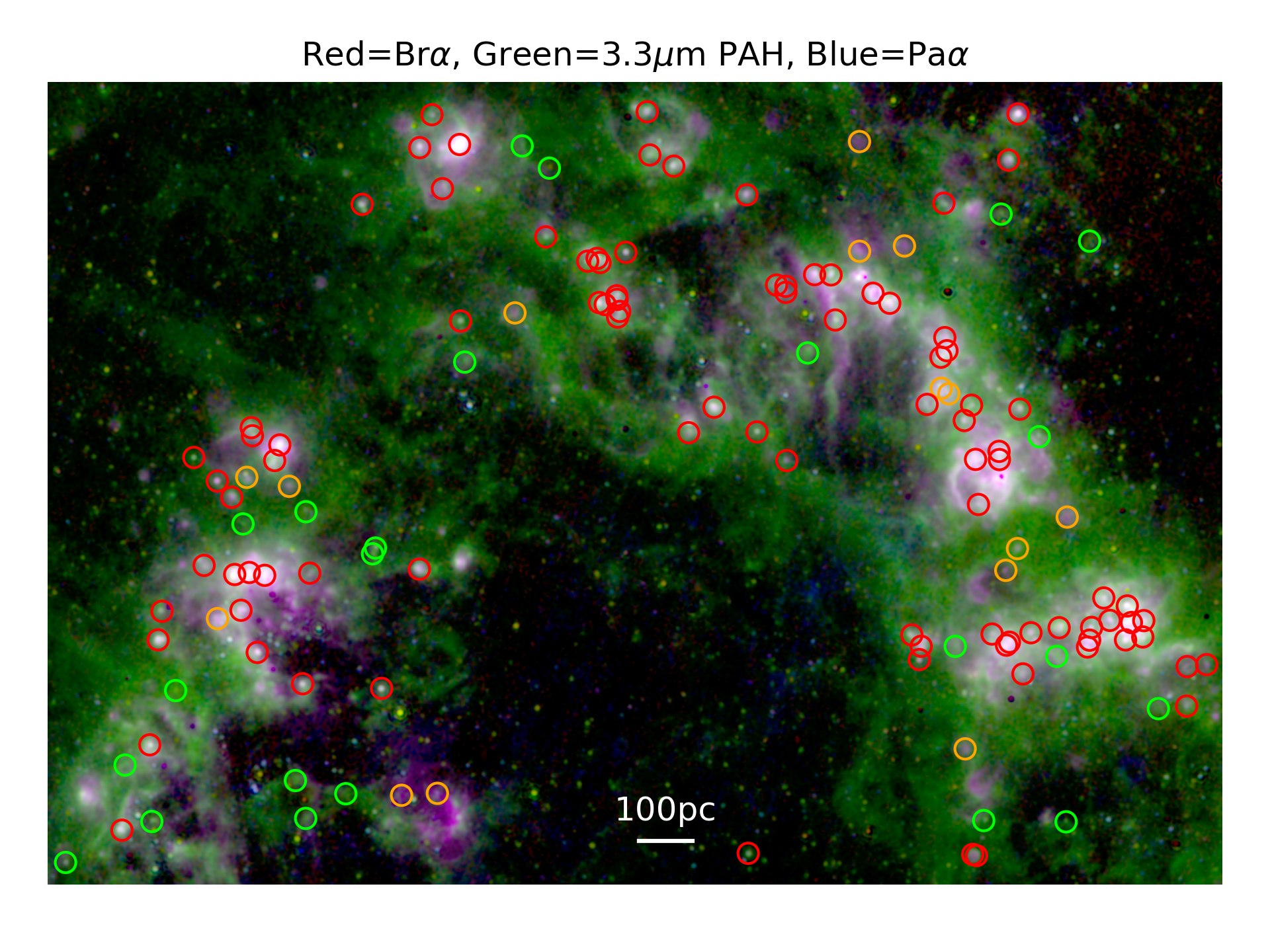}
\caption{ A three-color composite image showing the Br$\alpha$ (red channel), 3.3 $\mu$m PAH (green channel), and Pa$\alpha$ (blue channel) emission line maps for the same representative region as in Figure \ref{fig:f1}. Overlaid on top of the image are the 10 pixel (0.4$''$ or $\sim$19 pc) radius apertures used to measure the photometry of the three classes of sources. The eYSC--I sources (cospatial Pa$\alpha$, Br$\alpha$, and 3.3 $\mu$m PAH emission peaks) are shown in red, eYSC--II sources (hydrogen recombination line compact) are shown in orange, and PAH compact sources are shown in bright green. The white bar shows the 100 pc scale. }
\label{fig:f3}
\end{figure*}

Figure \ref{fig:f3} shows a three-color composite image of the Pa$\alpha$, Br$\alpha$, and 3.3 $\mu$m PAH emission line maps for the same representative region shown in the previous figures. Overplotted on the image are the 10 pixel radius apertures used to measure the photometry for the final source catalogs in this region. The three classes of sources correspond to the different colored circles. The eYSC--I sources (or cospatial peaks in Pa$\alpha$, Br$\alpha$, and 3.3 $\mu$m PAH emission) are shown in red, eYSC--II sources (hydrogen recombination line compact) are shown in orange, and PAH compact sources are shown in bright green. It is important to note that the catalogs displayed in Figure \ref{fig:f3} are not complete due to the removal of sources with peaks in 3.3 $\mu$m PAH and hydrogen recombination line emission that are greater than 4 pixels apart, but less than 20 pixels. The isolation criterion of 20 pixels is applied across the different classes to ensure that the eYSC--II and PAH compact sources are distinct for our measurements. As a result, there is no overlap between the different classes of sources in Figure \ref{fig:f3}. But within a given class, we do not require the same isolation criterion and the measurements may overlap, particularly in dense regions for eYSC--I; seen by the overlapping same colored circles in Figure \ref{fig:f3}. 

\hypertarget{3.3.1}{\subsubsection{Corrections}}

Our data is matched to the PSF of F444W, the lowest angular resolution filter, via convolution and thus we are sampling the same physical scales across wavelengths. In addition, our aperture size is much (about 6 times) larger than the FWHM of the PSF of F444W ($\sim$3.6 pixels) so we expect a correction for the aperture to be negligible and therefore do not apply one. However, typical star clusters may not be point sources in our images. The PSF FWHM of F444W is 0.145$''$ \citep{2023PASP..135d8001R} or ${\sim}$7 pc, compared to the average size of young, massive star clusters of ${\sim}$3 pc \citep{2017ApJ...841...92R}. As a result, we expect that the typical cluster may be slightly extended at the resolution of F444W. 

We measure aperture photometry around a sample of isolated, bright, slightly extended star cluster candidates in F444W for increasing aperture sizes ranging from 1 to 14 pixels in radius. The local background is measured in the same way as presented above, in annuli with an inner radius of 14 pixels and outer radius of 14$\sqrt{2}$ pixels, and is removed from the measurements. From these measurements, we determine that for an aperture size of 10 pixels in radius, the percent of the total flux of the sources recovered in F444W is ${\sim}$99$\%$, if the total flux is assumed to be reached at an aperture size of 14 pixels. In the case of eYSC--I, the centroid of the Pa$\alpha$ peak is used as the reference location where the photometry is measured. The matching criteria used to determine the cospatial peaks in the other emission lines leads to an offset in the measurements of those lines by at most 4 pixels. For the same isolated sources, we determine that when the source is offset from the aperture measurement by 4 pixels, the percent of the total flux of the sources recovered in F444W for 10 pixel radius apertures is ${\sim}$95$\%$, if we again assume that the total flux is reached at a size of 14 pixels. We do not apply a correction to our measurements to account for this as it depends on the offsets between the emission line peaks and is a relatively small effect, typically no more than a few percent. 

Another important consideration is that the line emission around young star clusters is more extended than the continuum emission. We make similar measurements as above for a small sample of the most isolated, slightly extended, line-selected candidate young star clusters (eYSC--I) and determine that for 10 pixel radius apertures, the percent of the total flux of the sources recovered in F444W is ${\sim}$96$\%$ when measured at the location of the peak in F444W and ${\sim}$89$\%$ when offset by 4 pixels. The decrease in the recovered flux here may be a result of the contribution of Br$\alpha$ to F444W. 

The HST/ACS F658N is contaminated by the [NII] lines at 6548 and 6583 \AA, and thus so is the H$\alpha$ emission line map. We assume that the average ratio between the [NII] lines and H$\alpha$ for NGC 628 is [NII]6548,6583/H$\alpha \, = \, 0.4$ \citep{2008ApJS..178..247K}, which is applied to the measured H$\alpha$ fluxes to provide a basic correction for [NII] contamination. Although it is a minor effect for our measurements, we also apply a correction to the measured fluxes to account for the transmission through the filters at the location of the redshifted emission lines, assuming z$\, = 0.00219$ for NGC 628 as listed in the NASA/IPAC Extragalactic Database (NED). 

\hypertarget{3.3.2}{\subsubsection{Luminosities, SFRs, and uncertainties}}

For each measurement, we derive emission line luminosity using $L\,$(erg$\,$s$^{-1})\,=\,$(3e-5$\, f_{\nu}\,/\lambda^{2}) \, BW  * 4 \pi d^{2}$, where $f_{\nu}$ is the flux density in Jy, $\lambda$ and $BW$ are the pivot wavelength and bandwidth of the filters in \AA, and $d$ is the distance to the galaxy in cm. For NIRCam filters, we use the pivot wavelengths (1.874, 3.365, and 4.055 $\mu$m) and bandwidths (240, 3470, 460 \AA) for F187N, F335M, and F405N, respectively, as given in the NIRCam documentation\footnote{\url{https://jwst-docs.stsci.edu/jwst-near-infrared-camera/NIRCam-instrumentation/NIRCam-filters}}. For HST/ACS/WFC F658N, we use the pivot wavelength listed in the FITS header (0.6584 $\mu$m) and bandwidth given in the ACS documentation (87.487 \AA)\footnote{\url{https://etc.stsci.edu/etcstatic/users_guide/appendix_b_acs.html}}. Emission line luminosity surface densities are calculated as $\Sigma_{L}\,$(erg$\,$s$^{-1}$$\,$pc$^{-2}$) = L$*$cos($i$)/A, where $i$ is the galaxy inclination angle and A is the physical area in pc$^{2}$ corresponding to the aperture measurement. For our apertures, the inclination corrected physical area is ${\sim}$1158 pc$^{2}$.

To estimate the uncertainties in the emission line luminosities, we perform a Monte Carlo calculation with the assumption that the photometric errors are normally distributed with a standard deviation given by their uncertainties. We assume an additional 10\% error on the distance to the galaxy to account for its uncertainty \citep[${\sim}6\%$;][]{2009AJ....138..332J,2021MNRAS.501.3621A} along with other unquantified sources of uncertainty, e.g. the continuum subtraction, calibration, etc. We simulate 10$^{4}$ random draws. The uncertainties in the emission line luminosities are derived as the standard deviation of the resulting distribution. Propagating the measurement uncertainties, we determine that the 3$\sigma$ detection limits for the emission line luminosities are 6.06, 1.85, 19.39, and 9.84 L$_{\odot}$ for Pa$\alpha$, Br$\alpha$, H$\alpha$, and the 3.3 $\mu$m PAH feature, respectively.

We estimate SFRs for our sources directly from the Br$\alpha$ luminosities. We use the Python code \texttt{PyNeb}\footnote{\url{https://pypi.org/project/PyNeb/}} \citep{2015A&A...573A..42L} to calculate the intrinsic H$\alpha$ to Br$\alpha$ luminosity ratio, assuming Case B recombination and typical HII region density n${\sim}\, 100$ cm$^{-3}$ and temperature T${\sim}\, 7000$ K for near solar metallicity. This gives $L_{H\alpha}$/$L_{Br\alpha}\sim 32$. From this intrinsic ratio and the H$\alpha$ calibration of \cite{2013seg..book..419C}, SFR$_{H\alpha}$ (M$_{\odot}$ yr$^{-1}$) = 5.5$\times$10$^{-42}$ L$_{H\alpha}$ (erg s$^{-1}$), we derive SFRs using SFR$_{Br\alpha}$ (M$_{\odot}$ yr$^{-1}$) = 1.76$\times$10$^{-40}$ L$_{Br\alpha}$ (erg s$^{-1}$). 

At this stage, we do not correct the Br$\alpha$ luminosity for dust extinction. As we discuss later in Sections \hyperlink{4}{4} and \hyperlink{5.3}{5.3}, the observed Pa$\alpha$ to Br$\alpha$ ratio is consistent with little-to-no extinction for our sources on average. However, this is somewhat complicated by the discrepancy that we observe between the Pa$\alpha$ to Br$\alpha$ and H$\alpha$ to Br$\alpha$ ratios. Yet, we find that both ratios lead to relatively low average color excesses, consistent with previous studies of the extinction in NGC 628 \citep[e.g.][]{2018ApJ...855..133K}. As a result, we expect the effect of extinction on Br$\alpha$ to be minor for our sources. Understanding the origin of the discrepancy and the impact of extinction in these sources will be important for future studies and it is being actively investigated (see Pedrini et al. in prep.).

Our measurements of the emission line luminosities will also be affected by the leakage of UV photons out of the ${\sim}19$ pc radius regions. We expect that between about 30$\%$ to 50$\%$ of the ionizing photons emitted by the local young star clusters will leak out of the HII regions \cite[e.g.][]{1997MNRAS.291..827O} and thus will be lost in terms of the measured ionized gas luminosity. These leaked photons are believed to be what powers the diffuse ionized gas (DIG). However, we expect that the nebular lines (H$\alpha$, Pa$\alpha$, and Br$\alpha$) will leak at the same fraction. For the 3.3 $\mu$m PAH emission, as long as the non-ionizing UV photons leak at a similar modality as the ionizing UV, we would expect that all the emission lines will be biased at roughly the same level. In this case, the leakage of UV photons would not have a large effect on our results. Currently, it is unclear if this is the case, and therefore the leakage of UV photons may affect our results. In the most extreme case that $1/2$ of the ionizing photons leak out of the HII regions and are not recovered in our apertures, we would expect the intrinsic SFRs to be a factor of ${\sim}$2 higher than our measurements from the observed Br$\alpha$ luminosities. 

\vspace{3mm}
\hypertarget{4}{\section{Results}}

\begin{figure}
\centering
\includegraphics[width=0.47\textwidth]{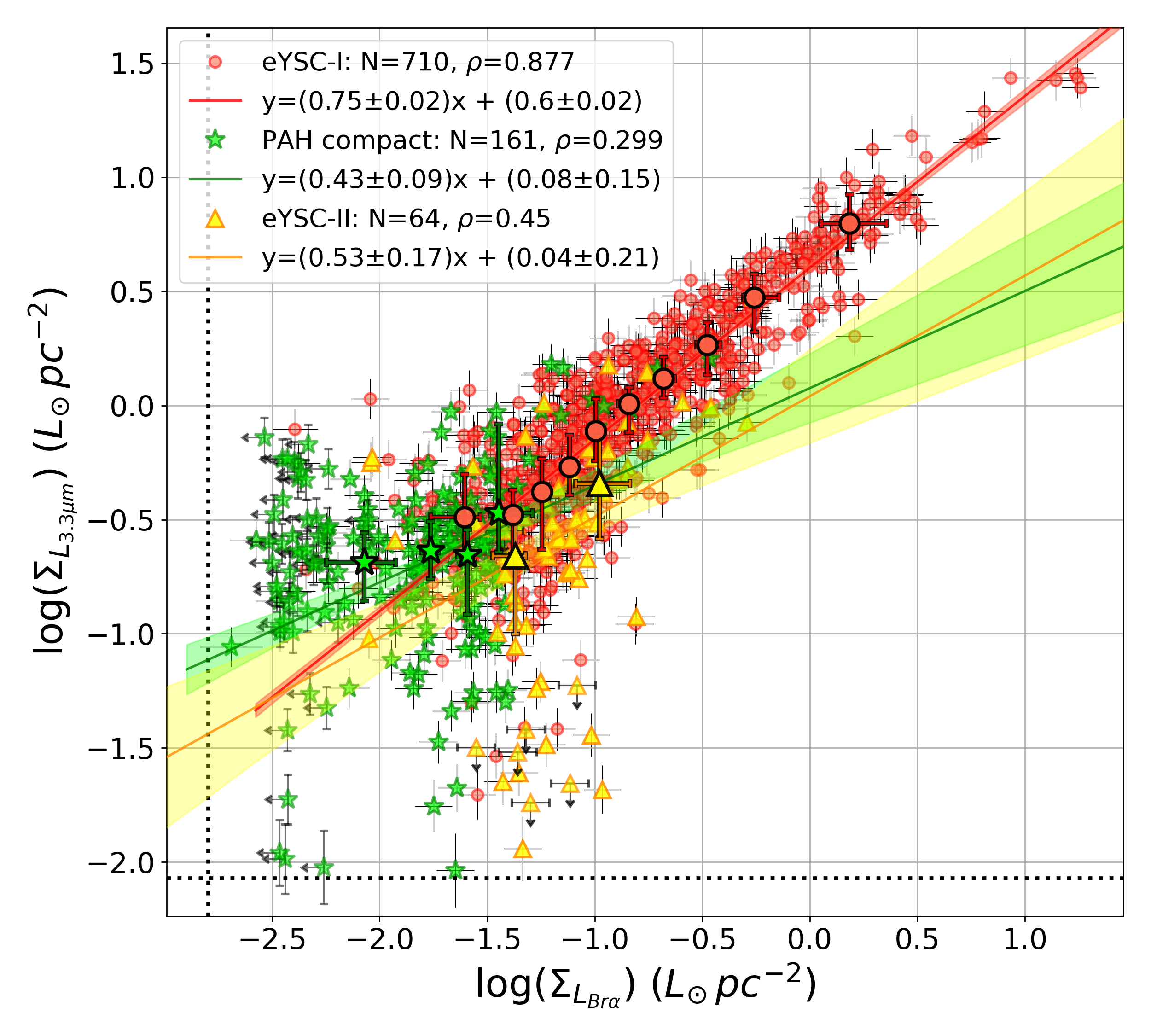}
\caption{ The 3.3 $\mu$m PAH luminosity surface density as a function of the Br$\alpha$ luminosity surface density for the three classes of sources. The eYSC--I sources (cospatial Pa$\alpha$, Br$\alpha$, and 3.3 $\mu$m PAH emission peaks) are shown as red circles, while eYSC--II sources (hydrogen recombination line compact) are shown by the yellow/orange triangles, and PAH compact sources are shown as green stars. Measurements with S/N $<$ 3 are shown as 3$\sigma$ upper limits. The colored lines show the best-fit relations determined from a Bayesian linear regression using the \texttt{LINMIX} package, while the shaded regions show the 1$\sigma$ confidence intervals. The large data points outlined in black show the median 3.3 $\mu$m PAH and Br$\alpha$ luminosity surface densities in equal size bins (n$\sim$71, 32, and 40; eYSC--I, eYSC--II, and PAH compact) of Br$\alpha$ luminosity, along with the range between the lower and upper quartiles shown by the error bars. The dotted lines show the 3$\sigma$ detection limits for our measurements based on the emission line map uncertainties. The figure caption gives the total number of sources (N), the Spearman correlation coefficient ($\rho$), and the values of the best-fit slope and y-intercept and their 1$\sigma$ uncertainties determined from the Bayesian regression for each class of sources. }
\label{fig:f4}
\end{figure}

Figure \ref{fig:f4} shows the 3.3 $\mu$m PAH versus Br$\alpha$ luminosity surface density for the three classes of sources. We determine the best-fit relations using the Bayesian linear regression implemented in the \texttt{LINMIX}\footnote{\url{https://github.com/jmeyers314/linmix}} Python code, which uses a linear mixture model algorithm developed by \cite{2007ApJ...665.1489K} to fit data with uncertainties on two variables. These are shown by the colored lines in Figure \ref{fig:f4}, which correspond to the mean best-fit slope and y-intercept of the traces. The colored regions show 1$\sigma$ confidence intervals given by the standard deviation of the best-fit parameters. For eYSC--I sources (cospatial Pa$\alpha$, Br$\alpha$, and 3.3 $\mu$m PAH emission peaks), we observe a tight relation between the 3.3 $\mu$m PAH and Br$\alpha$ emission, with a Spearman correlation coefficient ($\rho$) of 0.88. The best-fit for eYSC--I yields
\begin{eqnarray}
\label{eq:1}
    \text{log}\Big(\frac{\Sigma_{L_{3.3\mu m}}}{L_{\odot} \, pc^{-2}}\Big) = (0.75 \pm 0.02) \times \text{log}\Big(\frac{\Sigma_{L_{Br\alpha}}}{L_{\odot} \, pc^{-2}}\Big) \\ \nonumber
    + (0.60 \pm 0.02).
\end{eqnarray}
Therefore, we determine that the relation between these variables is sub-linear with a power-law exponent ($\alpha$) of 0.75. For eYSC--II (hydrogen recombination line compact) and PAH compact sources, the relation is much weaker with $\rho{\sim}\,$(0.45, 0.30) and $\alpha{\sim}\,$(0.53, 0.43), respectively. 

\begin{figure*}
\centering
\includegraphics[width=0.47\textwidth]{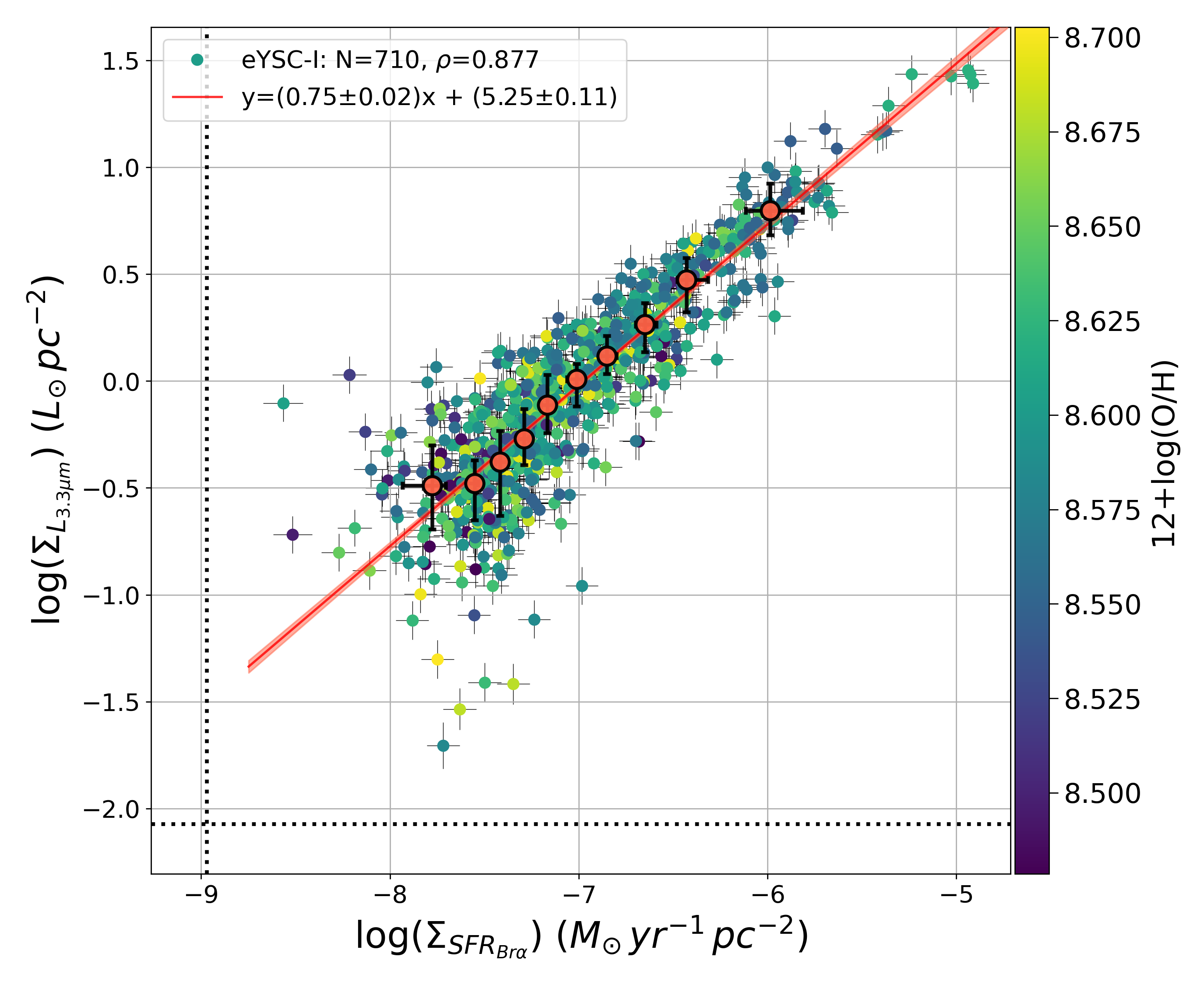}
\includegraphics[width=0.47\textwidth]{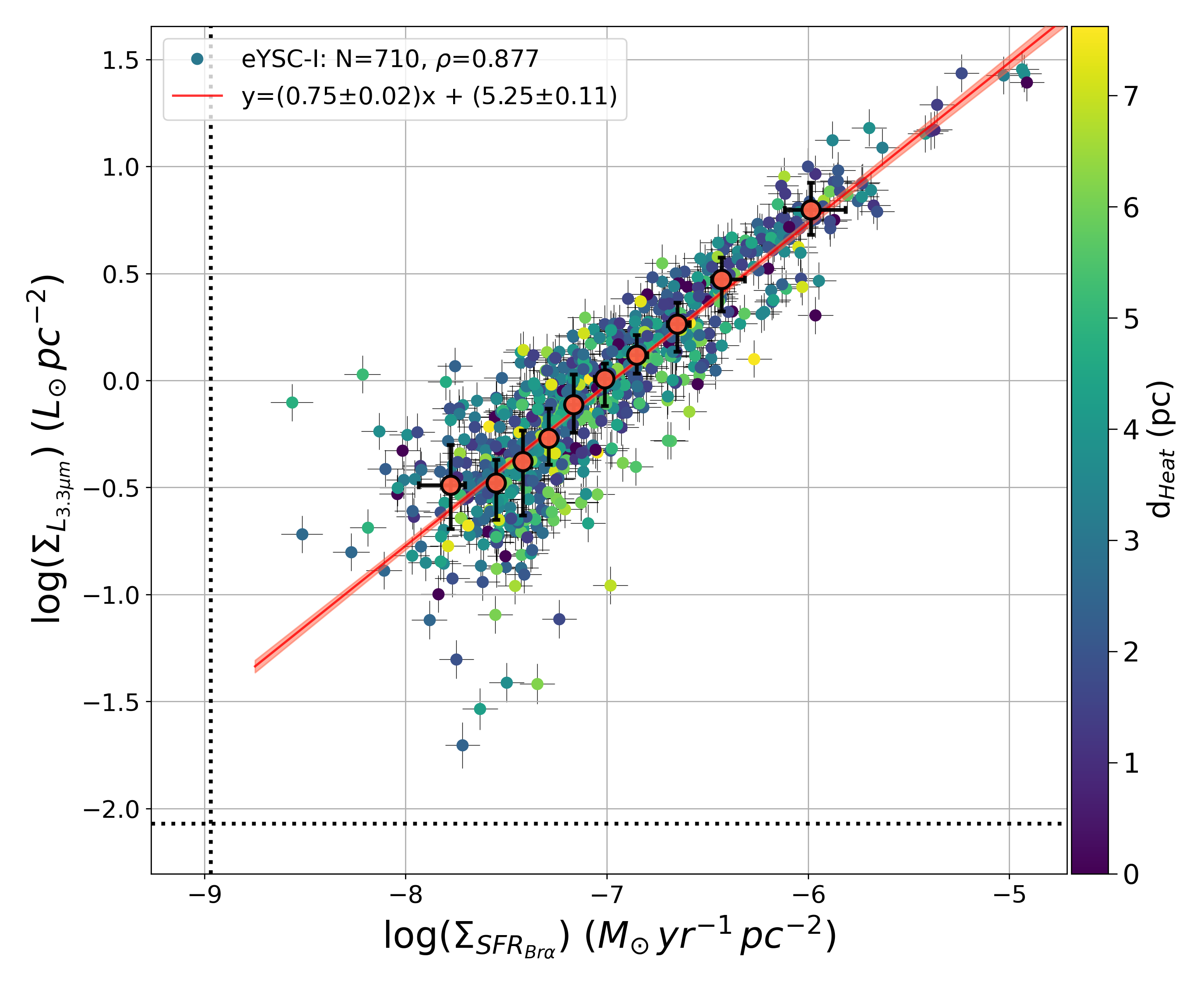}
\caption{ The 3.3 $\mu$m PAH luminosity surface density as a function of SFR surface density derived from Br$\alpha$ for eYSC--I sources. (Left panel) The points are color-coded by the oxygen abundance (gas-phase metallicity), derived from the galactocentric radius of each source in combination with the radial gradient of NGC 628 measured by \cite{2020ApJ...893...96B}. (Right panel) Points are color-coded by the distance (pc) between the local, young heating source (traced by the Pa$\alpha$ emission peak) and the nearest 3.3 $\mu$m PAH emission peak. See Figure \ref{fig:f4} for a more complete description.}
\label{fig:f5}
\end{figure*}

In Figure \ref{fig:f5}, we show the relationship between the surface densities of 3.3 $\mu$m PAH luminosity and SFR derived from Br$\alpha$ for eYSC--I sources. The red line shows the best-fit relation determined from the Bayesian linear regression. This corresponds to a new SFR calibration from the 3.3 ${\mu}m$ PAH emission given by: 
\begin{eqnarray}
\label{eq:2}
    \text{log}\Big(\frac{\Sigma_{SFR}}{M_\odot \, yr^{-1} \, pc^{-2}}\Big) = (1.33 \pm & 0.03) \times \text{log}\Big(\frac{\Sigma_{L_{3.3\mu m}}}{L_{\odot} \, pc^{-2}}\Big) \nonumber \\
    & - (6.97 \pm 0.15).
\end{eqnarray}
Expressed in terms of luminosity, the calibration is given by:
\begin{eqnarray}
\label{eq:3}
    \text{log}\Big(\frac{\mathrm{SFR}}{M_\odot \, yr^{-1}}\Big) = (1.33 \pm 0.03) \times \text{log}\Big(\frac{\mathrm{L}_{3.3\mu m}}{L_{\odot}}\Big) \nonumber \\ 
    - (7.98 \pm 0.08).
\end{eqnarray}
This calibration corresponds to ${\sim}$40 pc scales for tightly spatially connected, compact peaks in both ionized gas and PAH emission (eYSC--I). It remains unclear what regime this calibration applies to, outside of what we test in this study. The calibration coefficients are inevitably affected by the leakage of UV photons. Yet, we expect that only the y-intercept would be affected as the leakage of photons is not expected to be dependent on luminosity. In the case where the non-ionizing UV photons that heat the PAHs leak with a similar modality as the ionizing UV, the effect would be relatively minimal.

Although the relation between 3.3 $\mu$m PAH emission and SFR is relatively strong for these sources, there is also significant scatter that is likely too large to be accounted for by the measurement errors. The typical scatter in the data about the relation (Figure \ref{fig:f5}, Equation \ref{eq:2}) is determined to be about 0.14 dex, calculated as the mean orthogonal distance between the data and the best-fit relation. Several different sources may significantly contribute to the observed scatter, including variations in local ISM metallicity and PAH heating, both of which have been identified and studied for the PAH emission feature at 8 $\mu$m \citep[e.g.][]{2005ApJ...628L..29E,2007ApJ...666..870C,2007ApJ...656..770S,2008MNRAS.389..629B,2014ApJ...797..129L}. 

In the left panel of Figure \ref{fig:f5}, the data points are color-coded by the gas-phase metallicity as traced by the abundance of oxygen, which is derived from the galactocentric radius of each source in combination with the radial gradient of NGC 628 measured by \cite{2020ApJ...893...96B}. The radial metallicity gradient from \cite{2020ApJ...893...96B} is determined using temperature measurements from multiple auroral-line detections in individual HII regions of NGC 628. Our sources span a large range in galactocentric radius, from about 0.11$\arcmin$ to 2.98$\arcmin$ or about 0.31 kpc to 8.53 kpc. It is clear that there is no obvious relation between the scatter in Figure \ref{fig:f5} and differences in metallicity expected from the observed radial trend. 

In addition to differences in metallicity, variations in PAH heating may contribute to the observed scatter. For example, we expect that a PAH emission peak with a larger separation from the local ionized gas emission peak may have a larger contribution from heating by sources other than the local ionizing young star cluster. The right panel of Figure \ref{fig:f5} shows the data points color-coded by the distance (or offset) between the local heating source traced by ionized gas peak and the nearest 3.3 $\mu$m PAH emission peak (d$_{\text{Heat}}$), ranging from 0 to 4 pixels or about 0 to 7 pc. We find no indication of a strong correlation between the distance to the local PAH heating source and the observed scatter in the relation, at least for the offsets of our sources (${\lesssim}\,7$ pc). In Appendix \hyperlink{B}{B}, we present the results of a binning analysis in which our eYSC--I sources are divided into three statistically equal-size bins of metallicity and d$_{\text{Heat}}$.

\begin{figure}
\centering
\includegraphics[width=0.47\textwidth]{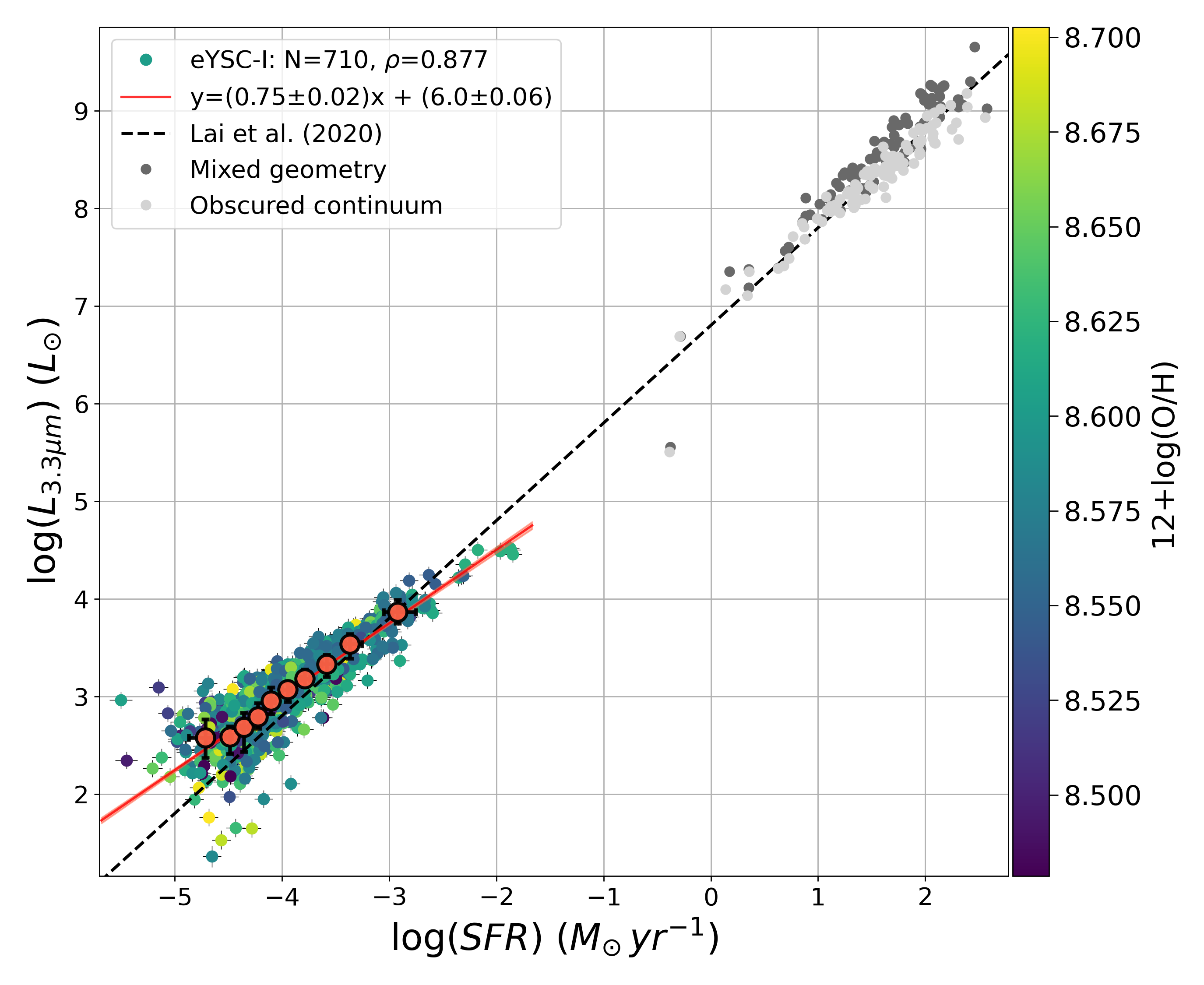}
\caption{The 3.3 $\mu$m PAH luminosity versus SFR. The grey points show measurements of a sample of local (0 $\leq$ z $\leq$ 0.2) galaxies from \cite{2020ApJ...905...55L} for two different assumptions on the dust geometry; mixed (dark grey) and obscured continuum (light grey). The dashed line shows the relation determined by \cite{2020ApJ...905...55L} (their Equation 1). See Figures \ref{fig:f4} and \ref{fig:f5} for a more complete description.}
\label{fig:f6}
\end{figure}

Figure \ref{fig:f6} shows a comparison of our results to the results of \cite{2020ApJ...905...55L}. \cite{2020ApJ...905...55L} is currently the only other study in the literature that investigates and calibrates the 3.3 $\mu$m PAH emission as an indicator of SFR. They analyze spectra from AKARI for a sample of nearby ($0\leq \,$z$\, \leq 0.2$) galaxies, selected to be PAH-bright. The grey data points in Figure \ref{fig:f6} show their galaxy-scale measurements under two different assumptions on the dust geometry. The dashed line shows their best-fit calibration, where SFRs are derived from a combination of [Ne II] and [Ne III]. There are clear differences in the slope determined by our study and by \cite{2020ApJ...905...55L}. They observe a slope consistent with one, while our data exhibits a substantially lower slope of 0.75.

\begin{figure*}
\centering
\includegraphics[width=0.9\textwidth]{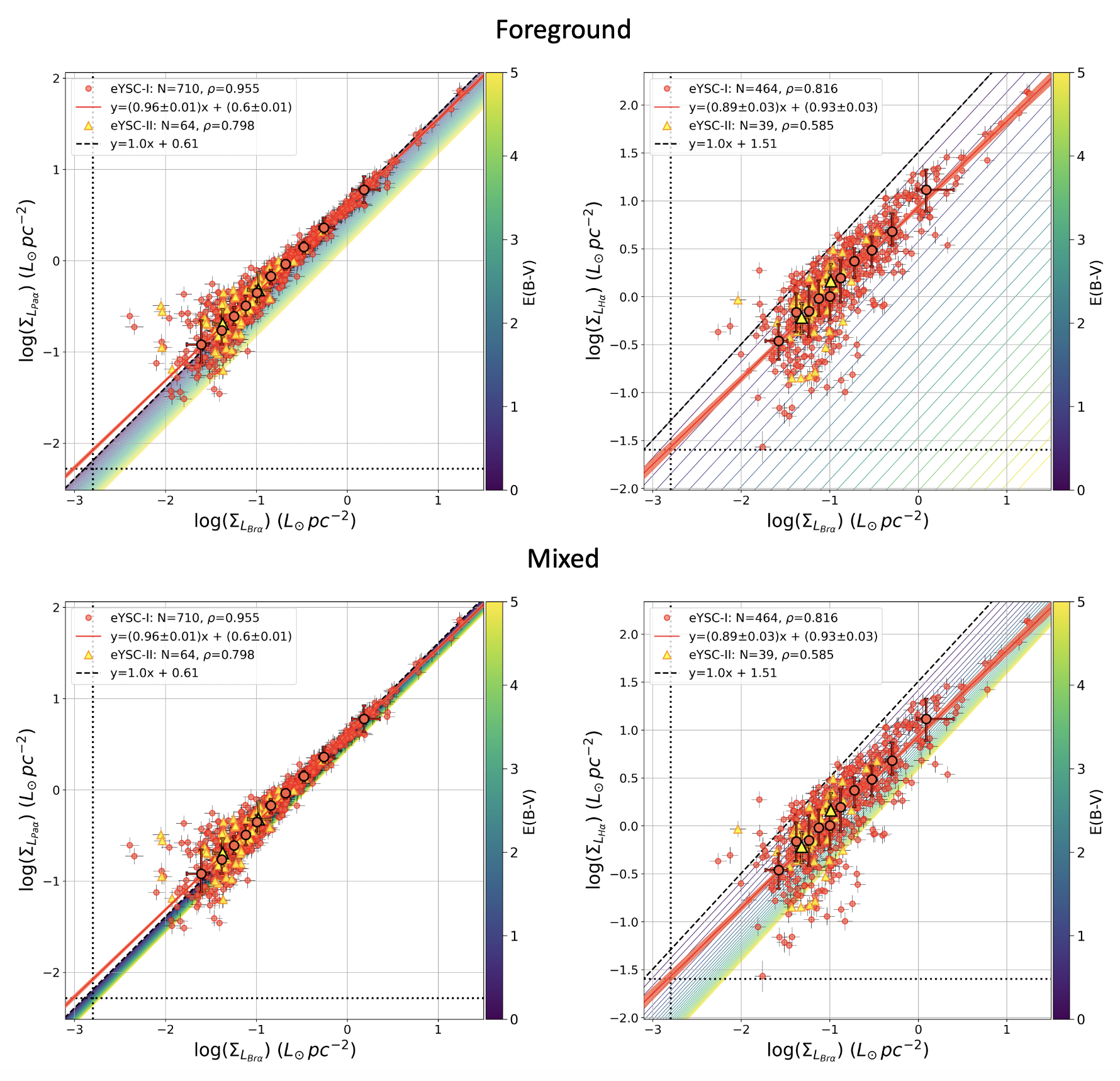}
\caption{The relations between the observed Pa$\alpha$ and Br$\alpha$ luminosity surface densities (left panels) and the H$\alpha$ and Br$\alpha$ luminosity surface densities (right panels) for eYSC--I (red circles) and eYSC--II (yellow/orange triangles). The dashed lines show the expected relation between the hydrogen recombination lines, given zero dust attenuation and an intrinsic ratio of Pa$\alpha$/Br$\alpha$ $\sim$ 4 or H$\alpha$/Br$\alpha$ $\sim$ 32 (n=100 cm$^{-3}$ and T=7000 K). The colored lines show how the expected relations change with increasing dust attenuation, measured by the color excess or E(B-V), for two different assumptions on the dust geometry: foreground (top panels) and mixed (bottom panels). We assume the Milky Way extinction curve of \cite{2021ApJ...916...33G}. See Figure \ref{fig:f4} for a more complete description. }
\label{fig:f7}
\end{figure*}

In Figure \ref{fig:f7}, we show the relations between the various hydrogen recombination lines for eYSC--I and eYSC--II sources. The left panels show the Pa$\alpha$ versus Br$\alpha$ luminosity surface density, while the right panels show the H$\alpha$ versus Br$\alpha$. The dashed lines correspond to the expected relations between the lines, given zero dust attenuation and an intrinsic ratio Pa$\alpha$/Br$\alpha$ $\sim$ 4 or H$\alpha$/Br$\alpha$ $\sim$ 32 (determined from \texttt{PyNeb} for Case B recombination, n=100 cm$^{-3}$, and T=7000 K). The colored lines demonstrate how the expected relations change with increasing dust attenuation, measured by the color excess or E(B-V) in steps of 0.2 mag, for two different assumptions on the dust geometry. We assume the models given in \cite{2021ApJ...913...37C}: (1) a foreground dust given by
\begin{displaymath}
L(\lambda)_{obs}=L(\lambda)_{int} 10^{[-0.4 E(B-V) k(\lambda)]} 
\end{displaymath}
 (top panels), where $L(\lambda)_{obs}$ and $L(\lambda)_{int}$ are the observed and intrinsic luminosities,  and (2) a homogeneous mixture of dust, stars, and gas given by
\begin{displaymath}
L(\lambda)_{obs}=\frac{L(\lambda)_{int}[1-e^{[-0.921 E(B-V) k(\lambda)]}]}{0.921 E(B-V) k(\lambda)}
\end{displaymath}
(bottom panels). We assume the Milky Way extinction curve $k(\lambda)$ determined by \cite{2021ApJ...916...33G}. 

As shown by the models, we expect increasing dust attenuation to lower the observed Pa$\alpha$/Br$\alpha$ and H$\alpha$/Br$\alpha$ ratios, thus decreasing the observed y-intercept of the relations in Figure \ref{fig:f7}. This is simply because the shorter wavelength line is more affected by increasing attenuation. However, we also expect dust attenuation to affect the slope, as the brightest sources will tend to be more extincted, leading to a decreased slope for higher average E(B-V). This is because sources with higher ionized gas luminosities generally have larger stellar masses and thus in the youngest stages are associated with larger giant molecular clouds (GMCs) and larger gas and dust masses and as a result, higher attenuation. 

We observe a tight relationship between the Pa$\alpha$ and Br$\alpha$ luminosities for our sources, with a correlation coefficient $\rho{\sim}\,$0.96 for eYSC--I (Figure \ref{fig:f7}, left panels). The relation between H$\alpha$ and Br$\alpha$ exhibits substantially more scatter with $\rho{\sim}\,$0.82 for eYSC--I. We find that the Pa$\alpha$ versus Br$\alpha$ relation is consistent with near zero attenuation for our sources (Figure \ref{fig:f7}, left panels). For eYSC--I, the slope or power-law exponent is measured to be $\alpha{\sim}$0.96$\pm 0.01$ and the y-intercept is $b{\sim}$0.60$\pm 0.01$, compared to the intrinsic values $\alpha$=$1.0$ and $b{\sim}$0.61. In contrast, we see clear evidence of the effect of extinction on the H$\alpha$ versus Br$\alpha$ relation (Figure \ref{fig:f7}, right panels). We observe a slope of $\alpha{\sim}$0.89$\pm 0.03$ and y-intercept of $b{\sim}$0.93$\pm 0.03$ for eYSC--I, compared to the intrinsic values $\alpha$=$1.0$ and $b{\sim}$1.51. Both the slope and y-intercept are significantly below (${\sim}$4$\sigma$ and ${\sim}$19$\sigma$, respectively) the expected values for zero attenuation. We determine the average E(B-V) to be about 0.5 mag for eYSC--I sources, from the H$\alpha$/Br$\alpha$ ratio and a foreground geometry. The relation between H$\alpha$ and Pa$\alpha$ is consistent with H$\alpha$ versus Br$\alpha$ for our sources.

\hypertarget{5}{\section{Discussion}}

\hypertarget{5.1}{\subsection{Sources of scatter in the 3.3 $\mu$m PAH calibration}}

As demonstrated in Figure \ref{fig:f5}, we find significant scatter (typically ${\sim}0.14$ dex) in the relation between 3.3 $\mu$m PAH emission and SFR traced by Br$\alpha$ for our young, embedded star cluster candidates. We discuss various dependencies and uncertainties identified for the PAH emission at 8 $\mu$m in Section \hyperlink{1}{1}, all of which may have some influence on the scatter we observe for the 3.3 $\mu$m PAH. Two of the most significant dependencies are on the local ISM metallicity and PAH heating. 

We determine that variations in the local ISM metallicity due to a global radial metallicity gradient do not account for the scatter observed in the relation between 3.3 $\mu$m PAH emission and SFR (Figures \ref{fig:f5} and \ref{fig:f8}, left panels). When binning into three statistically equal-size metallicity bins for eYSC--I, we find that the best-fit slopes and y-intercepts for each bin are consistent within ${\sim}3\sigma$ and that the correlation coefficients are nearly identical (see Appendix \hyperlink{B}{B}). The different bins of metallicity visually appear to occupy nearly the same space in Figure \ref{fig:f8}. These results are expected given the range in metallicities observed for our sources from about 12+log(O/H)${\sim}$8.5-8.7 (Figure \ref{fig:f5}, left panel), or moderately sub-solar to about solar metallicity. Typically, the 8 $\mu$m PAH emission has been observed to depend strongly on metallicity only at relatively low oxygen abundance, 12+log(O/H)$\lesssim$ 8.3 \citep[e.g.][]{2005ApJ...628L..29E,2007ApJ...666..870C}. Given that most of our sources in this galaxy are expected to have slightly sub-solar metallicity based on the observed radial gradient, it is unlikely that the dominant contributor to the scatter observed in the 3.3 $\mu$m PAH emission is due to metallicity variations in our sources.

One caveat here is that the observed radial metallicity gradient for NGC 628 derived by \cite{2020ApJ...893...96B} exhibits substantial intrinsic scatter, with a variation in oxygen abundance of up to 0.5 dex for fixed galactocentric radius. It is possible that more accurate metallicity determinations for our sources would reveal an underlying trend. To improve our understanding of the effect of varying ISM metallicity, it will be important to expand our sample of galaxies to include a larger range of metallicities, including both metal-poor systems like NGC 4449 \citep[average 12+log(O/H)$=$8.26;][]{2012ApJ...754...98B} and metal-rich systems like M83 \citep[characteristic 12+log(O/H)$=$8.73;][]{2016ApJ...830...64B}. We are currently in the process of analyzing NIRCam imaging from the FEAST program for both of these targets, which will be the subject of a future paper. It will also be enlightening to investigate individual systems like M101 that exhibit a highly diverse range of environments/metallicities in addition to a much steeper and tighter radial gradient \citep[see][]{2020ApJ...893...96B}.

In the right panel of Figure \ref{fig:f5}, we find no obvious correlation between the scatter in the 3.3 $\mu$m PAH versus SFR relation and the distance between the local PAH heating source, traced by the ionized gas peak, and the nearest 3.3 $\mu$m PAH peak (d$_{\text{Heat}}$). Separating the data into three equal-size bins of d$_{\text{Heat}}$, we determine that the best-fit slopes and y-intercepts for each bin are consistent within ${\sim}1\sigma$, but that there is a slight decrease in $\rho$, the correlation coefficient, towards larger d$_{\text{Heat}}$ (Appendix \hyperlink{B}{B}). We calculate that $\rho{\sim}0.92$ for eYSC--I sources with d$_{\text{Heat}}\leq2.17$ pc, while $\rho{\sim}0.81$ for d$_{\text{Heat}}>3.82$ pc (Table \hyperlink{t3}{3}). This suggests that some of the scatter observed in the relation may be due to the offset between the ionized gas and 3.3 $\mu$m PAH peaks. The slightly larger scatter observed for sources with larger offsets between the peaks may be due to measurement error. The photometry is measured at the location of the Pa$\alpha$ peaks so sources with larger offsets may have slightly underestimated 3.3 $\mu$m PAH flux, particularly if the 3.3 $\mu$m PAH peak is extended or bright, which can increase the scatter of the relation. We would also expect that this can lead to a lower derived slope for the sources with higher d$_{\text{Heat}}$. Yet, the slope measured for the low and high d$_{\text{Heat}}$ bins is consistent within ${\sim}1\sigma$, so the effect is relatively minor.

Variations in PAH heating may also play a role in increasing the scatter of the 3.3 $\mu$m PAH versus SFR relation shown in Figure \ref{fig:f5}. PAH emission peaks that have larger offsets from a local, young heating source may receive higher contributions from heating by the general radiation field. Yet, given that our measurements are local background subtracted and that the offsets between the two peaks are physically small for our sources, up to a maximum of only ${\sim}$7 pc, we expect that heating from the general non-ionizing UV radiation field will not have a large effect on our measurements. However, in dense regions in the galaxy, PAHs may be heated by multiple local sources of non-ionizing UV photons (star clusters). This can contribute to the scatter, particularly towards high luminosities, as sources in dense regions may exhibit elevated PAH flux relative to the ionized gas. 

There are a couple of other important sources of scatter that may contribute to our results shown in Figure \ref{fig:f5}. Most importantly, stochasticity in the stellar IMF likely dominates the contribution to the large scatter observed on the low mass/luminosity end. In Appendix \hyperlink{B}{B}, we present the results of a binning analysis in which we split the data for eYSC--I into three luminosity regimes. These correspond to: 1) high luminosity, above the expected H$\alpha$ luminosity of a 4 Myr old, 5000 M$_{\odot}$ star cluster, 2) low luminosity, below the expected H$\alpha$ luminosity of a 4 Myr, 1000 M$_{\odot}$ cluster, and 3) intermediate luminosity, between these expectations. Since stochasticity in the IMF is typically important for clusters with stellar mass below about 5000 M$_{\odot}$, we expect the high luminosity regime to be mostly free from the effects of stochastic sampling, while for the intermediate and low luminosity bins it likely becomes important. Around 80\% of our eYSC--I sources are expected to be in the regime where stochastic effects may be important (Figure \ref{fig:f9}). 

We determine that the scatter in the relation is much larger towards low luminosities, with a measured $\rho$ of 0.80, 0.62, and 0.38 for the high, intermediate, and low luminosity regimes respectively (Figure \ref{fig:f9} and Table \hyperlink{t3}{3}). The typical scatter, determined as the mean orthogonal distance between the data and the best-fit relation, is 0.09, 0.11, and 0.19 dex for the high, intermediate, and low luminosity regimes respectively. This large increase in the scatter towards low luminosity or mass is likely mostly due to stochasticity. At low mass, young star clusters will no longer fully sample the stellar IMF, particularly at the high mass end. Young clusters of the same mass may exhibit large differences in the production of ionizing photons as it depends sensitively on the high mass stars. This will increase the scatter in the ionized gas luminosity towards lower luminosities as stochastic sampling becomes more and more important. This provides a likely explanation of the trends in the scatter observed in Figure \ref{fig:f9}. Stochastic sampling is likely the dominant source of scatter in the relation at low luminosity. 

Another source of scatter is extinction. Neither the 3.3 $\mu$m PAH or Br$\alpha$ luminosity has been corrected for the effects of dust extinction. Yet, the results of Figure \ref{fig:f7} suggest that the E(B-V) of our sources is low on average, with little-to-no effect on the observed Pa$\alpha$/Br$\alpha$ ratios. This suggests that the effect due to differential attenuation between the 3.3 $\mu$m PAH emission and Br$\alpha$ is likely minor for our sources. We expect our measured calibration coefficients will be only slightly affected by extinction, likely no more than a few percent. 

Variations in the age of our sources may play a role in increasing the scatter. One reason to expect this is that at fixed cluster mass, we expect a decrease in the Br$\alpha$ luminosity with age. Theoretical models show that the ionized gas luminosity (e.g. H$\alpha$) depends on both the age and mass of the associated young star cluster. \cite{2021ApJ...909..121M} show the expectations for the Yggdrasil \citep{2011ApJ...740...13Z} single stellar population (SSP) models for the relation between the equivalent width of H$\alpha$ and Pa$\beta$ and the age of the associated young clusters, along with observational results from HST derived by SED fitting. They find both models and observations show a steep decrease in the equivalent width of H$\alpha$ and Pa$\beta$ with increasing age, reaching near zero by about 7-8 Myr. This correlation with age has also been studied in the past via the H$\alpha$ morphology, finding that more centrally concentrated H$\alpha$ morphologies are associated with younger clusters \citep[e.g.][]{2011ApJ...729...78W,2020ApJ...889..154W,2022MNRAS.512.1294H}. Our eYSC--I sources are generally young, embedded, and centrally peaked based on their selection, but they exhibit a range of ages between 1 and 6-7 Myr, based on SED fitting (Linden et al. in prep.). For the reasons given above, we expect that these variations in age may have the effect of increasing the observed scatter in the relation between 3.3 $\mu$m PAH emission and SFR.

Additionally, some small amount of scatter in the relation may be a result of overlapping measurements. Within a given class of sources (e.g. eYSC--I), we do not require the measurements to be completely isolated for our 10 pixel radius apertures. In dense regions where the emission peaks are tightly packed, our measurements show some overlap, seen as the intersecting same colored circles in Figure \ref{fig:f3}. This is most important for eYSC--I sources and may contribute some scatter, specifically towards high luminosity where the sources tend to reside in dense regions. However, we expect the effect to be relatively minor as it will not be important for sources that are near the same brightness, or when the offset is small since the brightest source will dominate. This effect is most significant in the case where there is another much brighter source near the edge of the aperture with significant differences in the spatial extent of the emission lines, e.g. more extended PAH emission. 


\hypertarget{5.2}{\subsection{The origin of the sub-linear relation}}

Figure \ref{fig:f6} shows that the measurements of \cite{2020ApJ...905...55L} are consistent with a linear relation between 3.3 $\mu$m PAH emission and SFR, yet our measurements show a clear deviation in slope, suggesting a sub-linear relation with a power-law exponent $\alpha{\sim}0.75$. The measurements of \cite{2020ApJ...905...55L} correspond to galaxy-integrated measures from AKARI spectra for a sample of nearby ($0\leq \,$z$\, \leq 0.2$) PAH-bright galaxies, consisting of mostly LIRGS/ULIRGS, and thus are quite different from our photometric measures of individual young star clusters and their associated HII regions at the scale of tens of parsecs. The galaxy sample of \cite{2020ApJ...905...55L} likely exists in the star formation dominated regime, where the majority of the non-ionizing UV photons that heat the PAHs are connected with recent star formation, rather than with a mix of young and old sources. Based on our selection of tightly spatially connected peaks in both ionized gas and PAH emission and the removal of the local background, we expect that this is also the case for our measurements. So what is the reason for the difference in slope? In the case of \cite{2020ApJ...905...55L}, the linear relation may simply be explained by the well-known luminosity-luminosity effect that is introduced from the distance squared proportionality of the luminosities. Linear relations among luminosities for galaxy samples with a large range of distances may be physically irrelevant.

Our results are consistent with a number of previous studies of resolved PAH emission in nearby galaxies. The PAH emission at 8 $\mu$m has been shown to exhibit a sub-linear relation with tracers of ionized gas \citep[e.g.][]{2005ApJ...633..871C,2007ApJ...666..870C}. \cite{2005ApJ...633..871C} find that the power-law exponent of the relation between 8 $\mu$m PAH emission and extinction-corrected Pa$\alpha$ is about 0.79 at 500 pc scales using Spitzer/IRAC and HST imaging of M51. More recently with JWST/NIRCam, \cite{2023ApJ...944L...9L} derive a power-law exponent of ${\sim}$0.6 for the relation between the 8 $\mu$m PAH emission and corrected H$\alpha$ at ${\sim}$80 pc scales in NGC 628. However, this study does not subtract the local background from the measurements and thus may be affected by PAH heating from the general UV radiation field. \cite{2023ApJ...957L..26L} use JWST/NIRSpec/IFU spectroscopy to study the 3.3 $\mu$m PAH emission on the scale of ${\sim}$200 pc in the starburst ring around the AGN of NGC 7469 (D${\sim}$70 Mpc). They study the relation between SFR derived from the 3.3 $\mu$m PAH and derived from the [NeII] and [NeIII] emission lines and find that SFR$_{3.3}$ is about 27$\%$ higher than SFR$_{Ne}$ on average. The relation between the 3.3 $\mu$m PAH and neon emission derived from their data could be consistent with a sub-linear relation, but it is unclear as the authors do not explicitly fit the relation, and their data also exhibits a very narrow range of luminosity. 

The emerging result is that the relation between PAH emission and SFR on small scales in galaxies is sub-linear. This suggests the presence of secondary processes that contribute to the relation. For the 3.3 $\mu$m PAH feature, the relation is well below linear (${\sim}12\sigma$) with a slope of ${\sim}0.75$ (Figure \ref{fig:f5}), suggesting that its use as a SFR indicator is complicated. The central question is then, what drives the sub-linear trend? For our sources, there are several possible origins for the observed sub-linear relation. One possibility is a result of variations in PAH heating, where a deviation from a slope of one could be explained by the increasing contribution of PAH heating by UV photons in the general radiation field at lower luminosity or SFR surface density. This could flatten the observed relation since sources with the lowest Br$\alpha$ luminosity will tend to be the ones with the largest contribution of heating by the diffuse radiation field, leading to increased PAH flux relative to the ionized gas towards low luminosities. Yet, this is unlikely to be the dominant driver of the sub-linear relation observed in our data as our measurements are local background subtracted. The removal of the local background from our measurements should mostly account for the PAH heating component by the general UV radiation field.

Another effect that can flatten the observed relation between PAH and ionized gas emission is stochasticity in the stellar IMF. As previously discussed in Section \hyperlink{5.1}{5.1}, we expect around 80$\%$ of our eYSC--I sources are in the regime where stochastic sampling may be important (see Figure \ref{fig:f9}). In the stochastic regime, young star clusters of the same mass may show substantial differences in the ionizing photon rate since it depends sensitively on the massive stars with mass M$\, >$ 15 M$_{\odot}$ that may not be fully sampled. Yet, PAHs are heated by non-ionizing UV photons, which will be relatively less affected. As a result, low mass, stochastically sampled clusters of similar mass may exhibit large differences in the ionized gas luminosity, while the PAH luminosity remains relatively unaffected. \cite{2011ApJ...741L..26F} find that stochastic sampling produces an asymmetric effect on the H$\alpha$/FUV ratio, where at low luminosity H$\alpha$/FUV decreases with the FUV luminosity. This is analogous to our ratio of Br$\alpha$ to 3.3 $\mu$m PAH emission. Similarly, in the stochastic regime, we may expect to see a decrease in the Br$\alpha$ to 3.3 $\mu$m PAH ratios with decreasing luminosity. This would have the effect of flattening the relation between PAH and ionized gas emission at low luminosity. 

In Appendix \hyperlink{B}{B}, we bin the data for eYSC--I into three luminosity regimes, based on the importance of stochastic sampling. We determine that the intermediate and high luminosity bins are consistent within ${\sim}1\sigma$ in terms of the derived slope and y-intercept, but that the low luminosity bin is over $4\sigma$ below, with a power-law exponent of $\alpha{\sim}0.41$ (Figure \ref{fig:f9} and Table \hyperlink{t3}{3}). This suggests that stochastic sampling has an important role in our sample in flattening the relation at low luminosity, consistent with our expectations. However, we note that the best-fit parameters are not significantly different (within ${\sim}1\sigma$) when fitting the full luminosity range or when fitting only the high luminosity regime, above the expected SFR of a 5000 M$_{\odot}$, 4 Myr old star cluster. This suggests that the inclusion of the low mass sources, where stochastic sampling is important, in our sample mostly affects the scatter and not the overall determined slope or intercept, which provides some justification for retaining these sources in the calibration given in Equation \ref{eq:2}.

Alternatively, the deviation in slope from unity could also be explained by the destruction of PAHs in more intense ionizing environments (i.e. a deficit in PAH emission at high Br$\alpha$ luminosity). This effect has been studied extensively in previous works finding that the abundance of PAHs decreases both for harder \citep[e.g.][]{2006A&A...446..877M,2014MNRAS.444..757K,2018MNRAS.481.5370M} and more intense radiation fields \citep[e.g.][]{2017ApJ...837..157S,2018ApJ...864..136B}. More recently, \cite{2023ApJ...944L..16E} use JWST/NIRCam imaging to show evidence for an anti-correlation between the PAH fraction and ionization parameter within HII regions in NGC 628. These results suggest that in more intense/harsh ionizing environments (high surface densities of Br$\alpha$ luminosity), we may expect to see a relative decrease in the PAH luminosity. However, it is unclear at what luminosity PAH destruction becomes important as it depends on the conditions of the ISM (e.g. metallicity). At high metallicity, the shielding provided by large grains may be effective at preventing the destruction of PAHs. Therefore, we may expect to see a transition or turnover in the relation towards high luminosity as shielding is overcome and PAH destruction becomes efficient.

Binning our eYSC--I sources into three luminosity regimes (Appendix \hyperlink{B}{B}), we determine that the intermediate and high luminosity bins are consistent within $1\sigma$ in terms of the derived slope and y-intercept (Figure \ref{fig:f9} and Table \hyperlink{t3}{3}). This suggests that there is no obvious turnover in the relation towards high luminosities. Although, there may be some indication of a slight deviation from the relation for a few of the brightest sources with log($\Sigma_{SFR_{Br\alpha}})>-5.0$ (Figure \ref{fig:f9}). As a result, this suggests that PAH destruction could be important for our sources either 1) across a large luminosity range, with no observed transition due to shielding, or 2) only at the very highest luminosities. Interestingly, we see an indication of a slightly higher slope for the highest metallicity sources (left panel, Figure \ref{fig:f8}). The high metallicity bin for eYSC--I exhibits ${\sim}3\sigma$ higher slope compared to the intermediate and low metallicity bins (Table \hyperlink{t3}{3}). The higher slope at high metallicity could be explained by more efficient shielding in these environments from the destruction of PAHs towards high luminosities. More data is needed to determine if this trend is real or if it is spurious and a result of the high scatter and relatively low number of sources in the bins (N=237).

With the remainder of our data, it will be interesting to test much higher SFR surface densities like the central starbursting regions of M83, which will help fill in the points with log($\Sigma_{SFR_{Br\alpha}})>-5.5$ in Figure \ref{fig:f9}. In these much more intense star-forming environments, we expect the signature of PAH destruction to be more significant. This will help us get a better handle on its effect. We will be able to test whether or not there is a transition at high luminosity as shielding is overcome, which if not, may suggest that PAH destruction is indeed important at intermediate luminosities and thus is key in driving our sub-linear relation. Also, it will be important to add in much lower metallicity environments like NGC 4449, where PAH destruction becomes significant at much lower luminosities. 

Variations in the age of our sources may also be important in driving the sub-linear relation. As discussed in Section \hyperlink{5.1}{5.1}, eYSC--I sources exhibit a range of ages between 1 and 6-7 Myr (Linden et al. in prep.), and we expect that at fixed cluster mass, older sources are associated with lower numbers of ionizing photons and lower Br$\alpha$ luminosity. On the other hand, PAHs are heated by non-ionizing UV photons and thus may be relatively unaffected by differences in age of a few million years. As a consequence, we may expect that older sources (low Br$\alpha$ luminosity) will typically have higher 3.3 $\mu$m PAH to Br$\alpha$ ratios. This would have the effect of flattening the relation between 3.3 $\mu$m PAH and Br$\alpha$ emission. Further investigation here is essential. This will be addressed in future work, where catalogs of both the ages and masses of our sources will begin to shed light on the potential effects of aging.

Another possibility is that for the brightest sources, there may be PAH emission excited by the local young star cluster that is missing from our fixed 10 pixel or ${\sim}19$ pc radius apertures. This could be the case as generally the diffuse PAH emission is found to be more extended than the ionized gas (see Pedrini et al. in prep.). Given that this missing PAH flux would be subtracted with the local background, this could lead to a sub-linear relation as the highest Br$\alpha$ luminosity sources would show the most underluminous measured 3.3 $\mu$m PAH emission. However, the brightest source in our eYSC--I sample has a SFR${\sim}$0.01 M$_{\odot}$ yr$^{-1}$ (Figure \ref{fig:f6}), giving a ionizing photon rate Q${\sim}1.35$e$+$51 s$^{-1}$ for the calibration of \cite{2013seg..book..419C}. Assuming a recombination coefficient $\alpha _{B}{\sim}3.5$e$-$13 cm$^3$ s$^{-1}$ for a density n${\sim}\, 100$ cm$^{-3}$ and temperature T${\sim}\, 7000$ K for Case B recombination from \cite{1995MNRAS.272...41S}, we calculate that the Str\"{o}mgren radius of our brightest source is R$_{s}{\sim}15$ pc. Therefore, the Str\"{o}mgren radius of the largest HII region in our sample is smaller than our ${\sim}19$ pc radius apertures. The study by \cite{2019ApJ...876...62C} finds that the PAH extent equals that of the HII regions. For these reasons and given that our sample is selected as compact peaks in both ionized gas and PAH emission, we expect that our apertures capture the vast majority of the PAH emission associated with the local cluster. It is unlikely that the observed sub-linear relation is due to an underestimate in the measured PAH emission at high luminosities due to our fixed apertures. 

Figure \ref{fig:f4} shows that compared to eYSC--I, the other classes of sources show clear evidence of an even flatter relation between 3.3 $\mu$m PAH and ionized gas emission (Br$\alpha$), with a power-law exponent $\alpha{\sim}0.43$ (PAH compact) and $\alpha{\sim}0.53$ (eYSC--II). These results are to be expected based on the discussion above and the selection of these sources. The PAH compact sources lack a compact, bright peak in ionized gas emission down to the detection limits. For these sources, the measured ionized gas emission is either low significance or more extended. Generally, this implies that these sources do not contain massive stars that ionize the gas. Yet, they exhibit a compact peak in PAH emission. Given that PAHs are heated by non-ionizing UV photons, this implies these sources could be clusters with a lot of stars with mass M$\, >$ 2-3 M$_{\odot}$ and M$\, <$ 15 M$_{\odot}$, but few with mass M$\, >$ 15 M$_{\odot}$ that dominate the production of ionizing photons. This can either be a consequence of age or mass. PAH compact sources are likely either low mass or old star clusters, such that they do not produce significant ionizing photons. This is in agreement with the results of Linden et al. (in prep.), who determine by fitting the optical/IR SEDs of our sources that PAH compact sources are indeed on average both older and less massive than eYSC--I. Therefore, the significantly decreased slope for PAH compact sources can be understood on the basis that they are generally less massive, and thus more stochastically sampled, and older. The slope determined for PAH compact sources is comparable to the slope measured for the low luminosity regime of eYSC--I ($\alpha{\sim}0.41$, Figure \ref{fig:f9}), suggesting that the masses of these may be similar on average and that the difference in the detection of ionized gas is likely stochastically driven.

On the other hand, the eYSC--II sources lack a compact, bright peak in 3.3 $\mu$m PAH emission. These sources are massive and young enough to produce significant ionizing photons, but the lack of PAH emission suggests that they may have already cleared their natal gas and dust. They are expected to be young, but generally older than eYSC--I. Inherently, we may expect both eYSC--II and PAH compact sources to be generally fainter than eYSC--I, but it is important to note that there is also a selection bias imposed by selecting only the ones that are isolated from a bright peak in the other emission line by at least 20 pixels. This criterion selects only fairly isolated sources for the PAH compact and eYSC--II catalogs, which tend to have lower luminosity and mass and thus be more stochastically sampled. Yet, this is necessary to ensure that these other classes are distinct from eYSC--I for our measurements. This effect likely accounts for some of the decrease in slope and increase in scatter seen for the PAH compact and eYSC--II sources compared to eYSC--I in Figure \ref{fig:f4}. For eYSC--II, the lower slope compared with eYSC--I may again be explained by these sources being both older and lower mass on average. However, the low number of eYSC--II sources and high scatter makes the determination of the slope inherently uncertain. It is also important to note that both eYSC--II and PAH compact sources are not well-described by the best-fit relations shown in Figure \ref{fig:f4} and show high scatter, likely a consequence of being dominated by stochastic sampling.

\hypertarget{5.3}{\subsection{Dust attenuation and geometry}}

In Figure \ref{fig:f7}, we observe that the relation between the Pa$\alpha$ and Br$\alpha$ luminosity for our sources is consistent with the expected, intrinsic relation, or with near zero attenuation. On the contrary, the H$\alpha$ to Br$\alpha$ ratio shows clear evidence of attenuation (Figure \ref{fig:f7}). There are a few potential explanations for this interesting discrepancy. 

One explanation is that there exists a miscalibration between the NIRCam filters. To obtain a better handle on the magnitude of the observed discrepancy, we measure photometry on the emission line maps in larger 15 pixel radius apertures, local background subtracted via annuli with inner radius of 30 pixels and outer radius of 35 pixels. From these measurements, we derive the E(B-V) of eYSC--I sources from the Pa$\alpha$/Br$\alpha$, H$\alpha$/Pa$\alpha$, and H$\alpha$/Br$\alpha$ ratios, assuming a foreground geometry, the extinction curve of \cite{2021ApJ...916...33G},  and the same physical conditions of the gas as before. The median E(B-V) for eYSC--I is determined from these estimates to be 0.29 mag for Pa$\alpha$/Br$\alpha$ and 0.48 mag for both H$\alpha$/Pa$\alpha$ and H$\alpha$/Br$\alpha$. Therefore, the average E(B-V) of much larger regions is still found to be inconsistent between the ratios of the recombination lines. We find that if the Br$\alpha$ luminosities are ${\sim}4\%$ higher relative to Pa$\alpha$ then the median E(B-V) estimates are near consistent between the three emission line ratios. At this time, it is possible that the NIRCam filter flux calibrations are uncertain by up to 4$\%$, especially for the narrowbands and between the short and long wavelength channels (e.g. F187N and F405N). However, the NIRCam flux calibrations we use in this study are the most updated at the time of the publication of this article, which have estimated flux calibration uncertainties of $\lesssim2\%$\footnote{\url{https://jwst-docs.stsci.edu/jwst-data-calibration-considerations/jwst-calibration-uncertainties}}. It is unclear at this stage, but improved calibrations may help to close this discrepancy between the ratios of hydrogen recombination lines.

Our continuum subtraction methods are also inherently uncertain. Specifically, Br$\alpha$ has two important components that contribute to the underlying continuum in star-forming regions, both stellar and hot dust emission. This contributes additional uncertainty to the subtractions. It is unclear how well the linear interpolation between F277W and F444W accounts for these different components, yet the F405N is very near in wavelength to F444W and therefore the subtraction will be dominated by the continuum in F444W. We note here that the results above for the discrepancy in E(B-V) are also observed when using the F300M and F444W to continuum subtract F405N. Our sources are selected as bright peaks in both Pa$\alpha$ and Br$\alpha$ and thus generally have high line-to-continuum ratios and relatively small uncertainties due to the subtraction. Nevertheless, it is possible that the continuum subtractions could contribute uncertainties in the measured luminosities of a few percent.

Alternatively, the discrepancy could be accounted for by the extinction curve. In this work, we assume the extinction curve of \cite{2021ApJ...916...33G} ($G21$). If the extinction curve k($\lambda$) was flattened between Pa$\alpha$ and Br$\alpha$ such that [k(Pa$\alpha$)/k(Br$\alpha$)] ${\sim}$ $0.6\,$[k(Pa$\alpha$)/k(Br$\alpha$)]$_{G21}$, than the E(B-V) estimates would be consistent between the H$\alpha$/Br$\alpha$ and Pa$\alpha$/Br$\alpha$ ratios. Currently, the extinction curve is not well constrained at these longer NIR wavelengths and it will take the new capabilities of JWST to determine whether the extinction curve is indeed more flat in the NIR than previous studies suggest. \cite{2023A&A...671L..14F} measure the NIR extinction law in 30 Doradus with new JWST/NIRCam imaging and find that although it is similar in slope with established Milky Way extinction curves between about 1 to 4 $\mu$m, there is evidence of a flattening of the curve at wavelengths $\gtrsim\,$4 $\mu$m. More data is needed though to determine if this result holds true in general and to establish the wavelength of this transition. 

Given the low average E(B-V) of ${\sim}$0.5 mag for eYSC--I, it is also possible that the effect of dust attenuation on the Pa$\alpha$/Br$\alpha$ ratio is small enough to be difficult to detect in this sample. This could be a source for the observed discrepancy. Yet for an E(B-V)${\sim}$0.5 mag, we would expect to measure a Pa$\alpha$/Br$\alpha$ ratio of ${\sim}$3.69 compared to the intrinsic value of ${\sim}$4.06, assuming a foreground geometry, Case B recombination, density n${\sim}\, 100$ cm$^{-3}$, and temperature T${\sim}\, 7000$ K. This corresponds to an expected y-intercept of b${\sim}$0.57$\pm0.01$ in the left panels of Figure \ref{fig:f7}. Thus the deviation from the intrinsic value of 0.61 would be detectable on the level of ${\sim}4\sigma$. As a result, even with the low average E(B-V), we would expect to detect the effect of dust attenuation on the Pa$\alpha$/Br$\alpha$ ratio in the majority of our sources. The origin of this discrepancy will continue to be investigated in future papers.

The geometry of the dust is a key consideration here as it greatly affects how the dust attenuates the emission. Figure \ref{fig:f7} shows the effect of increasing E(B-V) on the intrinsic relations between the recombination lines for two different dust geometries: foreground (top panels) and mixed (bottom panels). It is clear in Figure \ref{fig:f7} that both of these dust geometries are consistent with our derived trends, but a mixed geometry leads to higher derived E(B-V). In the case of the mixed geometry, we see a number of sources that exhibit too low Pa$\alpha$/Br$\alpha$ and H$\alpha$/Br$\alpha$ ratios to be consistent with this dust geometry model (bottom panels, Figure \ref{fig:f7}), seen by the points to the right of the highest E(B-V) trend. Yet, from these results, it is not possible to say which dust geometry model best fits our sources. 
The geometry of the dust likely evolves as the star cluster evolves.
We expect our eYSC--I sources to be on average a few million years of age and in the process of emerging, so the correct dust geometry likely lies somewhere between these two limiting cases.

New studies with JWST are unveiling the complexity of the relative distribution and morphology of stars, gas, and dust (including PAHs) in HII regions like the Orion nebula \cite[e.g.][]{2023arXiv230816733C,2023arXiv230816732H,2023arXiv231101163P,2023arXiv231008720P}. Given how complicated the geometries can be, orientation may be an important factor to consider as well. A blister HII region observed from different orientation angles will show variations in the relative attenuation of the star cluster and the emission lines. However, in our case the selection of eYSC--I sources as tightly spatially connected peaks in both ionized gas and PAH emission may help to mitigate any variations due to the relative geometries.

\hypertarget{6}{\section{Conclusions}}

In this paper, we present maps of ionized gas (Pa$\alpha$ and Br$\alpha$) and 3.3 $\mu$m PAH emission across NGC 628 created from new JWST/NIRCam observations from the FEAST survey. We discuss continuum subtraction techniques with the NIRCam bands, the selection of compact, young, embedded sources, the measurement of their PAH and ionized gas properties, and the calibration of the 3.3 $\mu$m PAH emission as a SFR indicator. Our main findings are the following:

\begin{itemize}
\item We find a tight (correlation coefficient $\rho$${\sim}$0.9) sub-linear (power-law exponent $\alpha$${\sim}$0.75) relation between the 3.3 $\mu$m PAH and Br$\alpha$ luminosities for our candidate emerging young star clusters (eYSC--I; cospatial peaks in both ionized gas and PAH emission) at ${\sim}40$ pc scales. From these measurements, we derive a novel SFR calibration from the 3.3 $\mu$m PAH luminosity, given in Equation \ref{eq:2}. The derived calibration coefficients may be affected by the leakage of UV photons. 

\item The scatter observed in the relation between 3.3 $\mu$m PAH emission and SFR traced by ionized gas is too large to be accounted for by the measurement errors. This scatter does not correlate well with differences in ISM metallicity expected given the observed radial metallicity gradient of NGC 628. The dominant sources of the scatter likely originate from variations in PAH heating, variations in the age of our sources, and stochastic sampling of the stellar IMF.  

\item The sub-linear relation between 3.3 $\mu$m PAH emission and SFR for our sources is likely explained by a combination of variations in age, PAH destruction in more intense ionizing environments, and stochasticity in the IMF at intermediate to low luminosities. This is supported in part by our binning analysis of eYSC--I and by our compact 3.3 $\mu$m PAH selected sources (PAH compact). PAH compact sources exhibit a lower power-law exponent ($\alpha$${\sim}$0.43), consistent with being on average less massive (and therefore more stochastically sampled) and/or older than eYSC--I. 

\item Correlating the hydrogen recombination lines, we find a tight relationship between the Pa$\alpha$ and Br$\alpha$ luminosities for eYSC--I, with a correlation coefficient $\rho{\sim}$0.96. The power-law exponent and y-intercept are consistent with the intrinsic values expected for zero dust attenuation. The relationship between the H$\alpha$ and Br$\alpha$ luminosities exhibits substantially more scatter with $\rho{\sim}$0.82 and a power-law exponent and y-intercept that are well below the intrinsic values. This discrepancy could be explained by a miscalibration between the NIRCam filters of up to 4\%. Alternatively, it may be due to errors in the continuum subtraction. The discrepancy could also be accounted for if the extinction curve is flatter in the NIR than recent pre-JWST Milky Way studies \citep[e.g.][]{2021ApJ...916...33G} suggest. Both a mixed and a foreground dust geometry are consistent with our measurements in all but some of the most extincted sources, where the mixed geometry can not account for the data. 
\end{itemize}

This paper demonstrates the ability of the 3.3 $\mu$m PAH emission feature observed by JWST/NIRCam to trace star formation in local systems on the fundamental scale of individual star clusters. However, the significantly sub-linear relation ($\alpha{\sim}0.75$) found on this scale with SFR traced by ionized gas emission suggests that the use of the 3.3 $\mu$m PAH emission as a SFR indicator is complicated. There are likely multiple secondary processes that contribute to the relation, some of which have been studied previously with the 8 $\mu$m PAH feature. Therefore, we suggest that caution is necessary when using the 3.3 $\mu$m PAH feature to trace star formation. The calibration of the 3.3 $\mu$m PAH feature has major applications to surveys of high redshift galaxies, where it can be used to push dust-obscured SFR estimates out to z${\sim}$7 with JWST/MIRI. Future studies will be required, both to further understand the 3.3 $\mu$m PAH emission and its dependency on local ISM environment, heating, etc., and to establish the connection between local and high redshift estimates where the different physical scales, star-forming environments, and inability to isolate the feature from the underlying stellar and dust continuum may significantly alter the results. 

\section*{Acknowledgments}
The authors would like to thank the anonymous referee for various thoughtful suggestions that helped further enhance this manuscript. 

This work is based in part on observations made with the NASA/ESA/CSA James Webb Space Telescope (JWST). The data were obtained from the Mikulski Archive for Space Telescopes (MAST) at the Space Telescope Science Institute (STScI), which is operated by the Association of Universities for Research in Astronomy, Inc., under NASA contract NAS 5-03127 for JWST. These observations are associated with program $\#$ 1783. Support for program $\#$ 1783 was provided by NASA through a grant from STScI. The specific observations analyzed can be accessed via \dataset[https://doi.org/10.17909/zcw1-6t85]{https://doi.org/10.17909/zcw1-6t85}. Support to MAST for these data is provided by the NASA Office of Space Science via grant NAG 5–7584 and by other grants and contracts.

The authors acknowledge the team of the `JWST-HST-VLT/MUSE-ALMA Treasury of Star Formation in Nearby Galaxies', led by coPIs Lee, Larson, Leroy, Sandstrom, Schinnerer, and Thilker, for developing the JWST observing program $\#$ 2107 with a zero-exclusive-access period.

This work is also based on observations made with the NASA/ESA Hubble Space Telescope, and obtained from the Hubble Legacy Archive, which is a collaboration between STScI/NASA, the Space Telescope European Coordinating Facility (ST-ECF/ESA) and the Canadian Astronomy Data Centre (CADC/NRC/CSA).

This research has made use of the NASA/IPAC Extragalactic Database (NED) which is operated by the Jet Propulsion Laboratory, California Institute of Technology, under contract with NASA.

BG acknowledges support from JWST GO 1783. AA and AP acknowledge support from the Swedish National Space Agency (SNSA) through the grant 2021- 00108. KG is supported by the Australian Research Council through the Discovery Early Career Researcher Award (DECRA) Fellowship (project number DE220100766) funded by the Australian Government and by the Australian Research Council Centre of Excellence for All Sky Astrophysics in 3 Dimensions (ASTRO~3D), through project number CE170100013. MRK is supported by the Australian Research Council through Laureate Fellowship FL220100020. 

\facilities{JWST/NIRCam, HST/ACS}

\software{astropy \citep{astropy:2013,astropy:2018}, photutils \citep{larry_bradley_2019_3568287}, SAOImageDS9 \citep{2003ASPC..295..489J}, SEP \citep{1996A&AS..117..393B,2016JOSS....1...58B}, PyNeb \citep{2015A&A...573A..42L}, LINMIX \citep{2007ApJ...665.1489K}}

\appendix
\vspace{-6mm}
\hypertarget{A}{\section{Variants for the continuum subtraction}}

With the inclusion of the additional NIRCam filters from the PHANGS--JWST program (ID 2107), presented in \cite{2023ApJ...944L..17L}, we are able to test various continuum subtraction techniques for the 3.3 $\mu$m PAH emission using combinations of the F277W, F300M, F360M, and F444W filters. Table \hyperlink{t2}{2} lists the results for the calibration of the relation between 3.3 $\mu$m PAH emission and SFR traced by Br$\alpha$ for a number of variants on the continuum subtractions. All the methods that use F444W as the long wavelength filter for subtraction are consistent in both the best-fit slope (or power-law exponent $\alpha$) and y-intercept ($b$) within ${\sim}$3$\sigma$. Using the F300M instead of the F277W as the short wavelength continuum filter (unscaled) gives an $\alpha$ and $b$ that are about 2$\sigma$ lower. Scaling the continuum image by a factor of 1.06 prior to the subtraction (using F277W) gives a slope and y-intercept that are about 1$\sigma$ higher than the unscaled version. Continuum subtraction methods that use the F360M as the long wavelength filter show the largest differences. These give the lowest determined slope, ${\sim}$5$\sigma$ below our adopted subtraction method from F277W and F444W with continuum scaled by 1.06, and a y-intercept ${\sim}$4-6$\sigma$ below. The correction implemented by \cite{2023ApJ...944L...7S} for the F300M and F360M subtraction method increases the y-intercept by about 2$\sigma$ but has no effect on the measured slope. 

The galaxy SED is well known to exhibit a minimum around 3 $\mu$m as the continuum emission transitions from stellar-dominated to dust-dominated. We expect the F277W filter to be mostly dominated by stellar continuum. Yet, in regions of star formation, we expect the F300M to be mostly dust continuum emission \citep[e.g.][]{2021ApJ...917....3D}, similar to the F444W. As a result, we expect that using the F277W and F444W for the continuum subtractions introduces more uncertainty than using the F300M and F444W. However, the absorption by water ice at 3.05 $\mu$m \citep[e.g.][]{2004ApJS..151...35G,2020ApJ...905...55L} presents an additional complication. This absorption feature, if present, would affect the F300M measurement more significantly than the F277W. We expect this absorption feature may be significant only for highly obscured regions. Our sources show relatively low obscuration with an average E(B-V) of ${\sim}$0.5 mag. It remains unclear if water ice absorption significantly affects the measurement of the continuum around 3 $\mu$m for our sources, and understanding this will require follow-up spectroscopic studies. 

For the first FEAST galaxy observed and presented in this work, NGC 628, we chose to use the F277W and F444W for the continuum subtractions, as the F300M mosaic from the PHANGS program does not cover our full mosaics. We aim to account for nonlinearity between F277W and F444W by introducing the continuum scaling factor of 1.06, determined visually to provide the optimal subtraction of stars in the field. We note here that using F277W rather than F300M has a relatively minor effect on our results as the two methods are consistent within about 2$\sigma$ (Table \hyperlink{t2}{2}). For the remainder of the FEAST targets, we have implemented a switch in the filter selection to replace the F277W with F300M, as we expect that it may provide a more accurate continuum subtraction of the 3.3 $\mu$m PAH and Br$\alpha$ emission lines for our sources. 

The lower slope and intercept determined in the case of using the F300M and F360M for the continuum subtraction can likely be explained by contamination in F360M. The NIRCam F360M filter is contaminated by not only the 3.3 $\mu$m PAH feature, but also by the 3.4 $\mu$m ``aliphatic", and 3.47 $\mu$m ``plateau" features. So we expect that using the F360M will tend to overestimate the continuum under the 3.3 $\mu$m PAH emission, leading to an underestimate of the strength of this feature in star-forming regions. Comparing the maps directly, we determine that the continuum subtracted F335M flux is a factor of ${\sim}$0.93 lower on average in star-forming regions when using the F300M and F360M compared to using the F300M and F444W, consistent with our expectation. This provides a likely explanation of why we observe the lowest slope and y-intercept in the case of the F335M continuum subtraction via a simple linear interpolation between F300M and F360M. 

The work by \cite{2023ApJ...944L...7S} provides a first-order correction for this effect. They introduce a new approach that utilizes the observed F335M/F300M and F360M/F300M colors and the relations between them in PAH-dominated and PAH-faint lines of sight to derive a corrected F360M/F300M color with which to predict the F335M continuum. Our results show that compared to the methods that use F444W, this new approach still gives a lower derived slope and y-intercept. Comparing the maps, we find that the continuum subtracted F335M flux is a factor of ${\sim}$1.5 higher on average in star-forming regions when using the F300M and F360M corrected via the method of \cite{2023ApJ...944L...7S} compared to using the F300M and F444W. Based on its location in the SED, width, and the fact that we determine the bright Br$\alpha$ line to only contribute at a level of ${\sim}3\%$ in star-forming regions, we expect that the F444W receives less contamination compared to the F360M and thus may provide a more accurate and less uncertain continuum subtraction for F335M. If this is the case, these results suggest that the F335M continuum subtraction method of \cite{2023ApJ...944L...7S} may overestimate the strength of the 3.3 $\mu$m PAH feature in star-forming regions. However, the differing contribution from the hot dust continuum in each filter may complicate this interpretation.

Spectroscopy is required to determine which of these methods of continuum subtraction is the most accurate and this will be investigated in future work. Our team has an accepted Cycle 2 JWST/NIRSpec/MSA program to get 1--5 $\mu$m spectroscopy for ${\sim}$100 eYSCs in NGC 628. This will be essential in establishing and understanding the many complex components of the SED in this regime and their effect on our measurements/calibrations. This will in turn enable us to accurately calculate the strength of the 3.3 $\mu$m PAH feature around these sources and compare to the maps derived from NIRCam to determine the optimal continuum subtraction method. 

\begin{center}
\begin{table}
\centering
\caption{\hypertarget{t2}{Continuum Subtraction Variants}}
$^{\mbox{\textit{*}}}$ $\text{log}\Big(\frac{\Sigma_{L_{3.3\mu m}}}{L_{\odot} \, pc^{-2}}\Big)  = \alpha \, \text{log}\Big(\frac{\Sigma_{SFR_{Br\alpha}}}{M_\odot \, yr^{-1} \, pc^{-2}}\Big) + b$ \\[1mm]
\begin{tabular}{ l | c c c c}
\hline
\hline
\rule{0pt}{4ex} Continuum Subtraction (CS) Method & N &  $\alpha$ & $b$ & $\rho$$\,^{\mbox{\textit{a}}}$\\[2mm]
\hline
\rule{0pt}{3ex} F335M$_{CS}$ from F277W \& F444W & 718 & $0.73 \pm 0.02$ & $5.14 \pm 0.11$ & 0.875\\
\rule{0pt}{3ex} F335M$_{CS}$ from F300M \& F444W & 600 & $ 0.69 \pm 0.02$ & $ 4.92 \pm 0.11$ & 0.881\\
\rule{0pt}{3ex} F335M$_{CS}$ and F405N$_{CS}$ from F300M \& F444W & 600 & $ 0.69 \pm 0.02$ & $ 4.91 \pm 0.11$ & 0.887\\
\rule{0pt}{3ex} F335M$_{CS}$ from F277W \& F444W scaled 1.06$^{\mbox{\textit{b}}}$ & 710 & $0.75 \pm 0.02$ & $5.25 \pm 0.11$ & 0.877\\
\rule{0pt}{3ex} F335M$_{CS}$ from F300M \& F444W scaled 1.06 & 597 & $ 0.74 \pm 0.02$ & $ 5.20 \pm 0.12$ & 0.879\\
\rule{0pt}{3ex} F335M$_{CS}$ from F300M \& F360M & 601 & $ 0.66 \pm 0.02$ & $ 4.64 \pm 0.11$ & 0.867\\
\rule{0pt}{3ex} F335M$_{CS}$ from F300M \& F360M corrected$^{\mbox{\textit{c}}}$ & 601 & $ 0.66 \pm 0.02$ & $ 4.85 \pm 0.11$ & 0.867\\[1mm]
\hline
\end{tabular}
\begin{flushleft} 
\rule{0pt}{3ex}
\currtabletypesize{\sc Note}--- \\
\rule{0pt}{4ex}
$^{\mbox{\textit{*}}}$ Best-fit parameters determined from the Bayesian linear regression for eYSC--I sources, as in Figure \ref{fig:f5}. Unless otherwise noted, the Br$\alpha$ luminosity is derived by continuum subtracting F405N via an interpolation between F277W and F444W. \\
$^{\mbox{\textit{a}}}$ The Spearman correlation coefficient $\rho$.\\
$^{\mbox{\textit{b}}}$ Our adopted method (Figure \ref{fig:f5}, Equation \ref{eq:2}). Scaled refers to multiplying the derived continuum image by a factor of 1.06 prior to subtraction. This factor is visually determined to give the optimal subtraction of stellar point sources.\\
$^{\mbox{\textit{c}}}$ Correction applied to account for the contribution of the 3.3 $\mu$m, 3.4 $\mu$m ``aliphatic", and 3.47 $\mu$m ``plateau" PAH features to the F360M filter, via the F335M/F300M and F360M/F300M colors as given in \cite{2023ApJ...944L...7S} (their Equation 11). 
\end{flushleft}
\end{table}
\end{center}

\hypertarget{B}{\section{Binning analysis}}

\begin{figure*}[h]
\centering
\includegraphics[width=0.48\textwidth]{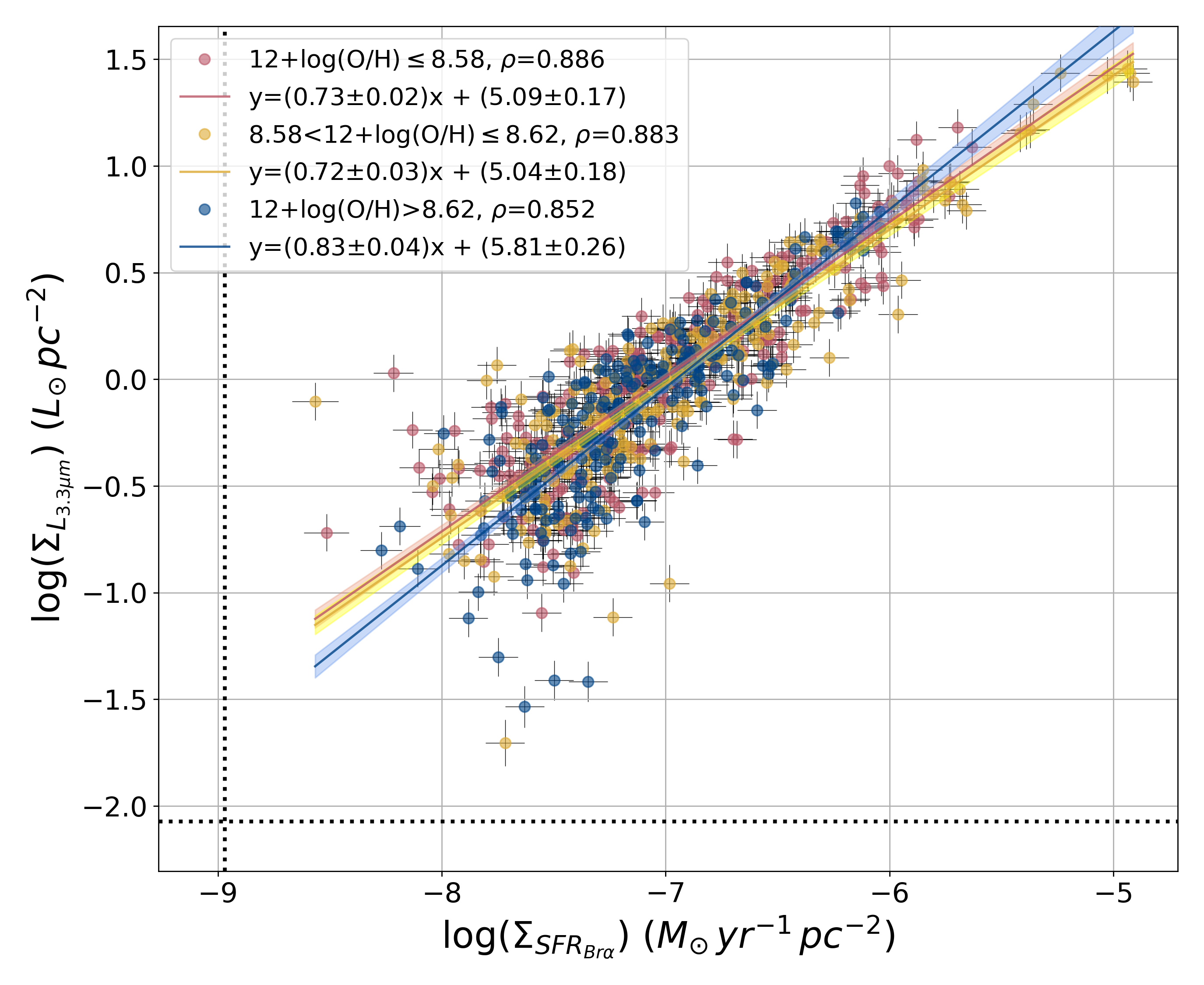}
\includegraphics[width=0.48\textwidth]{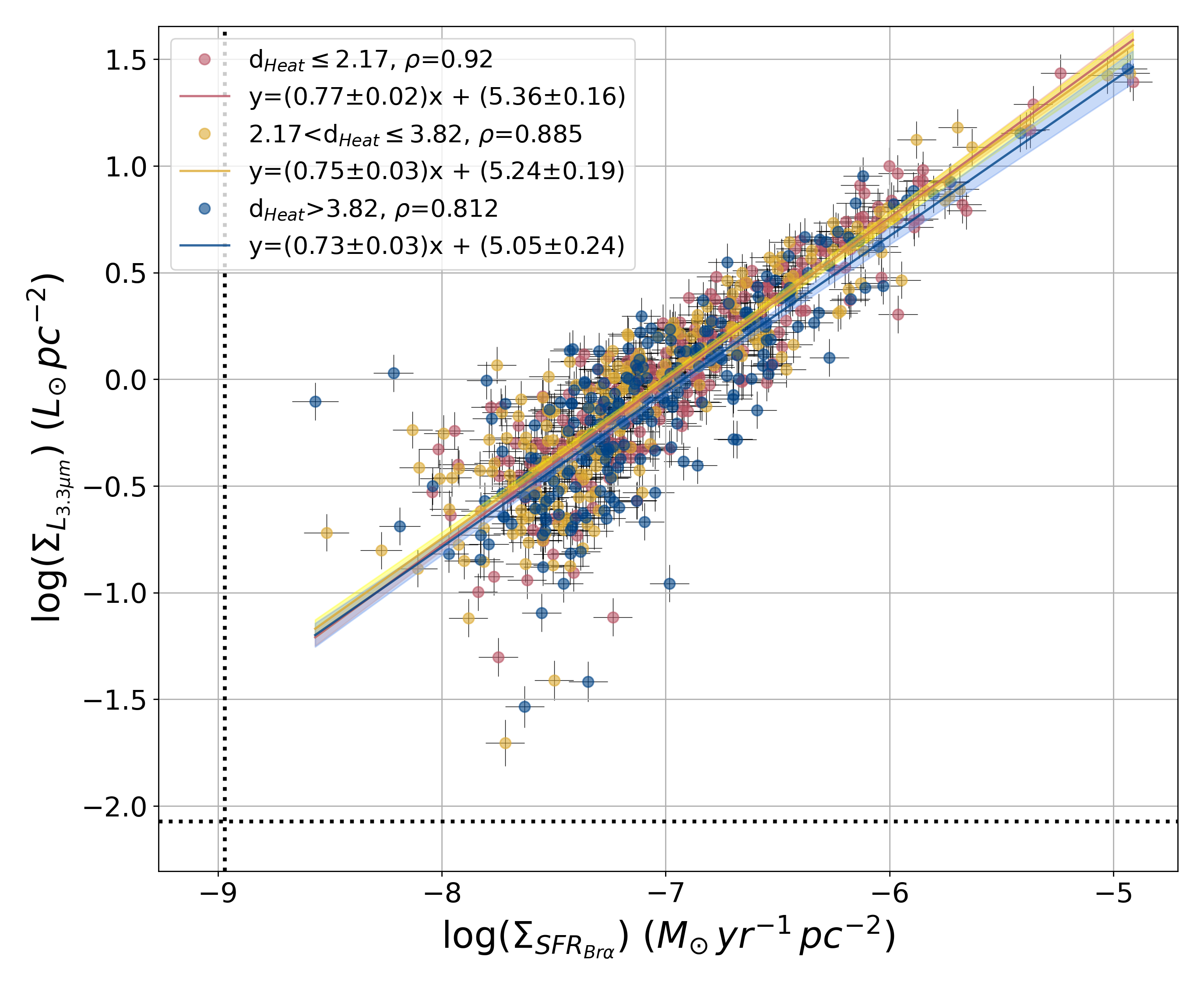}
\caption{The 3.3 $\mu$m PAH luminosity surface density versus SFR surface density derived from Br$\alpha$ for eYSC--I sources. (Left panel) The data is binned into three statistically equal size metallicity bins (N=237): low 12+log(O/H)$\leq$8.58 (red points), intermediate 8.58$<\,$12+log(O/H)$\,\leq$8.62 (yellow), and high 12+log(O/H)$>$8.62 (blue). The oxygen abundance is derived from the radial gradient of NGC 628 measured by \cite{2020ApJ...893...96B}. The colored lines show the best-fit relations for each bin determined by Bayesian regression. The dotted lines show the 3$\sigma$ detection limits. The caption gives the Spearman correlation coefficient ($\rho$) and the values of the best-fit slope and y-intercept and their 1$\sigma$ uncertainties for each bin. (Right panel) Binned into three equal size bins of d$_{\text{Heat}}$, the distance (pc) between the local, young heating source (traced by the Pa$\alpha$ peak) and the nearest 3.3 $\mu$m PAH emission peak: low d$_{\text{Heat}}\leq2.17$ pc (red), intermediate 2.17$<\,$d$_{\text{Heat}}\leq3.82$ pc (yellow), and high d$_{\text{Heat}}>3.82$ pc (blue).}
\label{fig:f8}
\end{figure*}

In Figure \ref{fig:f8}, we show the relation between the surface density of 3.3 $\mu$m PAH luminosity and SFR for eYSC--I, split into three statistically equal-size bins in metallicity (left panel) and d$_{\text{Heat}}$, the distance between the local heating source traced by the Pa$\alpha$ peak and the 3.3 $\mu$m PAH peak (right panel). The metallicity is derived from the radial oxygen abundance gradient of NGC 628 measured by \cite{2020ApJ...893...96B}. We fit the data in each bin independently with the Bayesian linear regression methods presented in Section \hyperlink{4}{4}. Table \hyperlink{t3}{3} lists the total number of sources (N), the best-fit power-law exponent ($\alpha$), y-intercept ($b$), and the Spearman correlation coefficient ($\rho$) for each of the bins in metallicity and d$_{\text{Heat}}$. There are no major differences in the best-fit parameters for the different bins of metallicity or d$_{\text{Heat}}$. The high metallicity bin has a slightly higher $\alpha$ and $b$, by ${\sim}3\sigma$, compared to the low and intermediate bins, but is near identical in terms of the scatter traced by $\rho$. For d$_{\text{Heat}}$, the bins are consistent within ${\sim}1\sigma$ in terms of $\alpha$ and $b$, but there is a minor decrease in $\rho$ towards larger d$_{\text{Heat}}$.

\begin{figure}[h]
\centering
\includegraphics[width=0.48\textwidth]{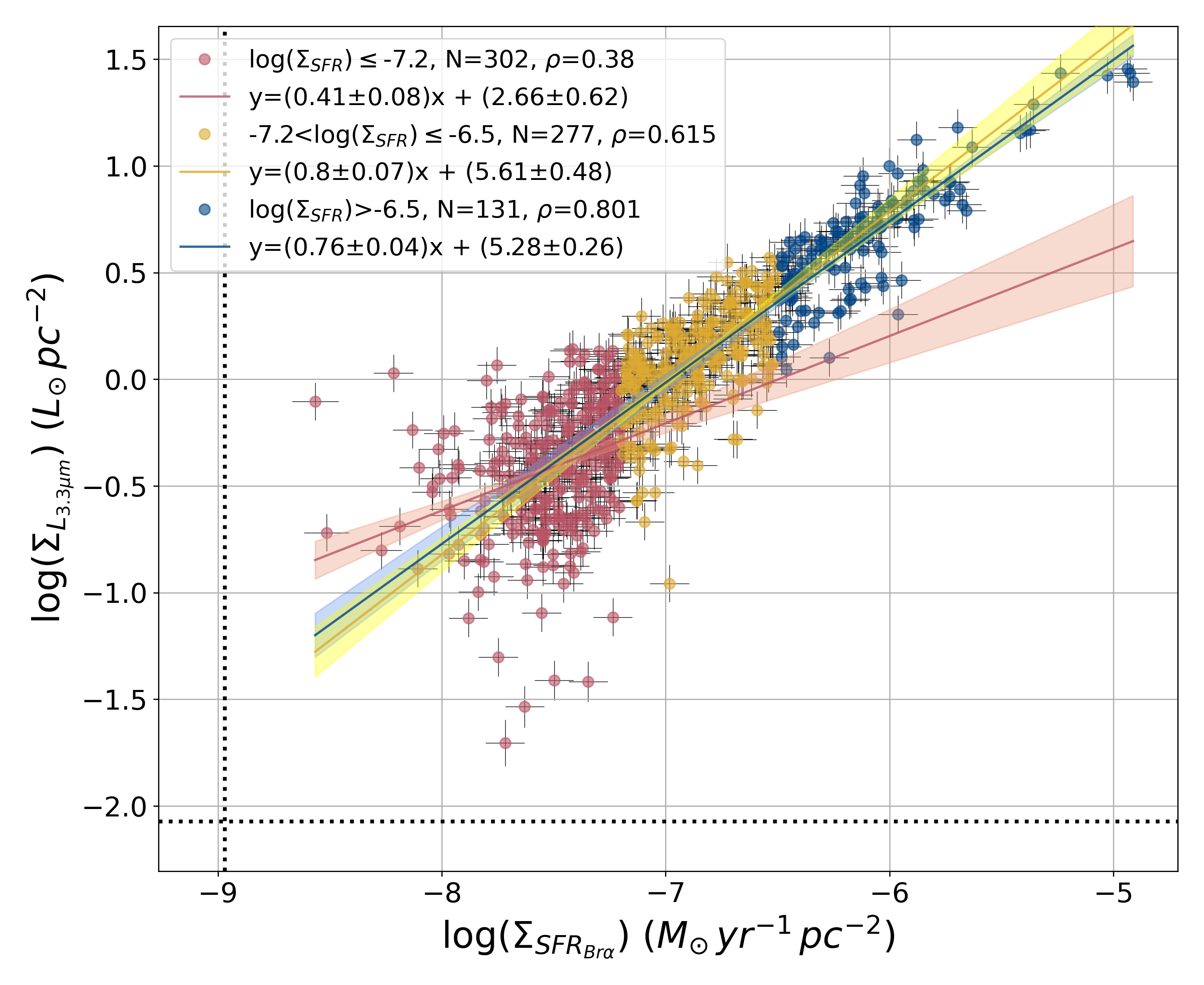}
\caption{The 3.3 $\mu$m PAH luminosity surface density versus SFR surface density for eYSC--I, binned into three luminosity regimes: low log($\Sigma_{SFR_{Br\alpha}})\leq-7.2$ (red points), intermediate $-7.2<\,$log($\Sigma_{SFR_{Br\alpha}})\leq-6.5$ (yellow), and high log($\Sigma_{SFR_{Br\alpha}})>-6.5$ (blue). The bin limits represent the log($\Sigma_{SFR}$) corresponding to the expected H$\alpha$ luminosity of a 4 Myr old cluster with a stellar mass of 5000 M$_{\odot}$ ($-$6.5) and 1000 M$_{\odot}$ ($-$7.2), based on Starburst99 models with Z=0.02 and the Padova AGB evolutionary tracks. See Figure \ref{fig:f8} for a more complete description. }
\label{fig:f9}
\end{figure}

 Based on their luminosities, our eYSC--I sources are expected to have a wide range of stellar masses. We expect the low luminosity/mass regime to be affected by stochastic sampling of the stellar IMF. Stochasticity in the IMF is typically thought to be important for star clusters with stellar mass below about 5000 M$_{\odot}$. Figure \ref{fig:f9} shows the 3.3 $\mu$m PAH luminosity surface density versus SFR surface density for eYSC--I, binned into three luminosity regimes. These regimes are determined in the following way. We estimate the expected H$\alpha$ luminosity of young (4 Myr) star clusters with a stellar mass of 5000 and 1000 M$_{\odot}$ from the Starburst99 models \citep{1999ApJS..123....3L} with a metallicity Z=0.02 and the Padova AGB stellar evolutionary tracks. This gives log(L$_{H\alpha}\,/$ erg s$^{-1}$)$\,=[37.78, 37.08]$ for $[5000, 1000]$ M$_{\odot}$, respectively. From the H$\alpha$ luminosity, we determine the SFR using the calibration of \cite{2013seg..book..419C}. We then divide by the same physical area as our measurements to determine log($\Sigma_{SFR}\,/$ M$_{\odot}\,$yr$^{-1}\,$pc$^{-2}$)$\,=[-6.5, -7.2]$ for the $[5000, 1000]$ M$_{\odot}$ cluster, respectively. These values of $\Sigma_{SFR}$ are used as limits to define the three bins. We expect the upper or high luminosity (or $\Sigma_{SFR}$) bin, above the expected SFR of a 5000 M$_{\odot}$, 4 Myr cluster, to be mostly unaffected by stochasticity. In the intermediate bin, stochasticity begins to be important and may affect our measurements, while the low luminosity bin is well within the stochastic regime. 

The best-fit relations for each of the three luminosity (or $\Sigma_{SFR}$) bins are shown in Figure \ref{fig:f9}. Table \hyperlink{t3}{3} lists the best-fit $\alpha$, $b$, and $\rho$ for each bin. The intermediate and high luminosity bins are consistent within $1\sigma$ in terms of $\alpha$ and $b$. The low luminosity bin shows a much lower slope and intercept, over $4\sigma$ below the other bins. We find an overall increase in the scatter of the trend towards lower luminosities, with a decrease in the measured $\rho$ from 0.80 to 0.38 for the high to low luminosity bins. These results suggest that stochastic sampling of the stellar IMF plays an important role in our sample in the lower luminosity/mass regime, both at increasing scatter and flattening the relation between 3.3 $\mu$m PAH emission and SFR.

\begin{center}
\begin{table}
\centering
\caption{\hypertarget{t3}{Binning Analysis}}
$^{\mbox{\textit{*}}}$ $\text{log}\Big(\frac{\Sigma_{L_{3.3\mu m}}}{L_{\odot} \, pc^{-2}}\Big)  = \alpha \, \text{log}\Big(\frac{\Sigma_{SFR_{Br\alpha}}}{M_\odot \, yr^{-1} \, pc^{-2}}\Big) + b$ \\[1mm]
\begin{tabular}{ l | c c c c }
\hline
\hline
\rule{0pt}{4ex}
\hspace{1.6cm} Bin & N & $\alpha$ & $b$ & $\rho$$\,^{\mbox{\textit{a}}}$ \\[2mm]
\hline
\rule{0pt}{3ex}
12+log(O/H)$\,\leq\,$8.58 & 237 & $ 0.73 \pm 0.02$ & $5.09 \pm 0.17$ & 0.886\\
\rule{0pt}{3ex} 8.58$\,<\,$ 12+log(O/H) $\,\leq\,$8.62  & 236 & $ 0.72 \pm 0.03$ & $ 5.04 \pm 0.18$ & 0.883\\
\rule{0pt}{3ex} 12+log(O/H)$\,>\,$8.62  & 237 & $ 0.83 \pm 0.04$ & $ 5.81 \pm 0.26$ & 0.852 \\[1mm]
\hline
\rule{0pt}{3ex} d$_{\text{Heat}}\leq2.17$ pc  & 237 & $ 0.77 \pm 0.02$ & $ 5.36 \pm 0.16$ & 0.920\\
\rule{0pt}{3ex} 2.17$<$ d$_{\text{Heat}}$ $\leq3.82$ pc  & 236 & $ 0.75 \pm 0.03$ & $ 5.24 \pm 0.19$ & 0.885\\
\rule{0pt}{3ex} d$_{\text{Heat}}>3.82$ pc  & 237 & $ 0.73 \pm 0.03$ & $ 5.05 \pm 0.24$ & 0.812\\[1mm]
\hline
\rule{0pt}{3ex} log($\Sigma_{SFR_{Br\alpha}})\leq-7.2$$^{\mbox{\textit{c}}}$  & 302 & $ 0.41 \pm 0.08$ & $ 2.66 \pm 0.62$ & 0.380\\
\rule{0pt}{3ex} $-7.2<\,$log($\Sigma_{SFR_{Br\alpha}})\leq-6.5$  & 277 & $ 0.80 \pm 0.07$ & $ 5.61 \pm 0.48$ & 0.615\\
\rule{0pt}{3ex} log($\Sigma_{SFR_{Br\alpha}})>-6.5$  & 131 & $ 0.76 \pm 0.04$ & $ 5.28 \pm 0.26$ & 0.801\\[1mm]
\hline
\end{tabular}
\begin{flushleft} 
\rule{0pt}{3ex}
\currtabletypesize{\sc Note}--- \\
\rule{0pt}{4ex}
$^{\mbox{\textit{*}}}$ Best-fit parameters determined from the Bayesian linear regression for eYSC--I sources, see Figures \ref{fig:f8} and \ref{fig:f9}. \\
$^{\mbox{\textit{a}}}$ The Spearman correlation coefficient $\rho$.\\
$^{\mbox{\textit{c}}}$ The bin limits are determined as the values expected from a 4 Myr old cluster with a stellar mass of 5000 M$_{\odot}$ ($-$6.5) and 1000 M$_{\odot}$ ($-$7.2), based on Starburst99 models with Z=0.02 and the Padova AGB evolutionary tracks.\\
\end{flushleft}
\end{table}
\end{center}

\clearpage
\bibliographystyle{aasjournal}
\bibliography{main.bib}

\end{document}